\documentclass[10pt]{amsart}
\usepackage{graphicx,dsfont}
\usepackage{subfigure} 
\usepackage{amsmath, amsfonts, amsthm, amssymb,verbatim} 
\usepackage[dvips,twoside]{geometry}

\usepackage{xr}
\usepackage[ps2pdf, bookmarks=true,
            bookmarksopen=true,colorlinks=false]{hyperref}
            
\newtheorem{theorem}{Theorem}[section]

\newtheorem{lemma}[theorem]{Lemma}

\numberwithin{equation}{section}

\newcommand \alphat {\widetilde \alpha} 

\newcommand \what {\widehat w} 
\newcommand \Xt {\widetilde X} 
\newcommand \Pt {\widetilde P} 
\newcommand \Qt {\widetilde Q} 
\newcommand \Et {\widetilde E}

\newcommand \bz {\begin{itemize}}
\newcommand \ez {\end{itemize}}
\newcommand \ben {\begin{enumerate}}
\newcommand\een {\end{enumerate}} 

\newcommand \R {\mathbb R} 
\newcommand{\Eqref}[1]{\eqref{#1}}
\newcommand{\Eqsref}[1]{\eqref{#1}}
\newcommand{\Sectionref}[1]{Section~\ref{#1}}  

\newcommand{\Lemref}[1]{Lemma~\ref{#1}}
\newcommand{\Propref}[1]{Proposition~\ref{#1}}
\newcommand{\Theoremref}[1]{Theorem~\ref{#1}}

\newcommand{\Figref}[1]{Figure~\ref{#1}}
\newcommand \auth {} 
\newcommand \jou {\textit}

\newcommand \muh M 
\newcommand \vv u

\newcommand \RR 		{\mathbb{R}}  
\newcommand \del 	\partial
\newcommand \eps 	\epsilon
\newcommand \lam 	\lambda 
\newcommand \be 		{\begin{equation}}
\newcommand\ee 		{\end{equation}}

\let\oldmarginpar\marginpar
\renewcommand\marginpar[1]{\-\oldmarginpar[\raggedleft\footnotesize #1]%
{\raggedright\footnotesize #1}}

\begin{document}

\title[Second-order hyperbolic Fuchsian systems.]{Second-order hyperbolic Fuchsian systems.
\\ 
Gowdy spacetimes and the Fuchsian numerical algorithm} 

\author[Florian Beyer and Philippe G. L{\tiny e}Floch]{Florian Beyer$^1$ and Philippe G. L{\scriptsize e}Floch$^2$ 
}

\date{}

\maketitle 

\footnotetext[1]{
Department of Mathematics and Statistics, University of Otago, P.O.~Box 56, Dunedin 9054, New Zealand, E-mail: {fbeyer@maths.otago.ac.nz}.
\newline
$^2$Laboratoire Jacques-Louis Lions \& Centre National de la Recherche Scientifique, Universit\'e Pierre et Marie Curie (Paris 6), 4 Place Jussieu, 75252 Paris, France. Blog: philippelefloch.wordpress.com
\newline 
\newline
Written in June 2010. Updated in Nov. 2010. \bf
A shortened version of the material presented in this 
preprint is included in:
{\sl F. Beyer and P.G. LeFloch, Second-order hyperbolic Fuchsian systems and applications,
Class. Quantum Grav. 27  (2010), 245012.}   
}

\begin{abstract} 
  This is the second part of a series devoted to the singular
  initial value problem for second-order hyperbolic Fuchsian
  systems. In the first part, we defined and investigated this general class of systems, and 
we  established a well-posedness theory in weighted Sobolev spaces.
This theory is applied here to the vacuum {\sl Einstein equations for Gowdy spacetimes} admitting, by definition,  
two Killing fields satisfying certain geometric conditions. 
We recover, by more direct and simpler arguments, the well-posedness results established earlier  
by Rendall and collaborators. In addition, in this paper we introduce a natural
  approximation scheme, which we refer to as the {\sl Fuchsian numerical algorithm} and is directly 
 motivated by our general theory. 
This algorithm provides highly accurate, numerical approximations of the solution to
the singular initial value problem. 
In particular, for the class of Gowdy spacetimes under consideration, various 
numerical experiments are presented which show the interest and efficiency of the proposed method.
Finally, as an application, we numerically construct Gowdy spacetimes containing a smooth, incomplete, 
  non-compact Cauchy horizon. 
\end{abstract}


\section{Introduction}

This is the second part of a series \cite{BeyerLeFloch1,BeyerLeFloch3} devoted to the initial
value problem associated with the Einstein equations for spacetimes endowed with symmetries.  
We are typically interested in spacetimes with Gowdy symmetry, and in 
formulating the Einstein equations with data on a singular hypersurface, 
where curvature generically blows-up. In the first part \cite{BeyerLeFloch1}, we introduced a class of partial differential equations, referred as as {\sl second-order Fuchsian systems,} and we established a general well-posedness theory within Sobolev spaces with weight (on the coordinate singularity). In the present paper, we tackle the treatment of actual models derived from the Einstein equations when suitable symmetry assumptions and gauge choices are made. 
  
We consider here $(3+1)$-dimensional, vacuum spacetimes $(M,g)$ with spacelike slices diffeomorphic to the torus 
$T^3$, satisfying the vacuum Einstein equations and the Gowdy symmetry assumption. That is, we assume the 
existence of an Abelian $T^2$-isometry group with spacelike orbits
and vanishing twist constants \cite{Chrusciel}. These so-called Gowdy spacetimes on $T^3$ were first
studied in \cite{Gowdy73}. In the past years, a combination of theoretical and numerical
works led to a detailed understanding of the Gowdy spacetimes, achieved 
by analyzing solutions to the Einstein equations as the singular boundary is approached; 
cf.~\cite{Chrusciel, IsenbergMoncrief, Moncrief, Ringstrom2, Ringstrom4, Ringstrom6}. 

Before we can indicate precisely our contribution in the present
paper, let us provide some background on Gowdy spacetimes. Introduce
coordinates $(t,x,y,z)$ such that $(x,y,z)$ describe spatial sections
diffeomorphic to $T^3$ while $t$ is a timelike variable.  We can
arrange that the Killing fields associated with the Gowdy symmetry
coincides with the coordinate vector fields $\del_y$,
$\del_z$, in a global manner, so that the spacetime metric reads
\begin{equation}
  \label{eq:gowdymetric}
  g=\frac 1{\sqrt t} \, e^{\Lambda/2}(-dt^2+dx^2)+t \, (e^P(dy+Qdz)^2+e^{-P}dz), \qquad t >0. 
\end{equation}
Hence, the metric depends upon three coefficients $P=P(t,x)$, $Q=Q(t,x)$,
and $\Lambda=\Lambda(t,x)$. We also assume spatial periodicity with
periodicity domain $U:=[0,2\pi)$.

In the chosen gauge, the Einstein's
vacuum equations imply the following second-order wave equations for $P,Q,$ 
\begin{equation}
  \label{eq:originalgowdy}
  \aligned
    P_{tt} + \frac{P_t}{t} - P_{xx} & = e^{2P} ( Q^2_t - Q^2_x),
    \\
    Q_{tt} + \frac{Q_t}{t} - Q_{xx} & = -2(P_t Q_t - P_x Q_x),
  \endaligned 
\end{equation}
which are decoupled from the wave equation satisfied by the third coefficient $\Lambda$:  
\begin{equation}
  \label{eq:evolLambda}
  \Lambda_{tt}-\Lambda_{xx}
  = P_x^2-P_t^2+e^{2P}(Q_x^2-Q_t^2).
\end{equation}
Moreover, the Einstein equations also imply constraint equations, which read 
\begin{subequations}
  \label{eq:constraints}
  \begin{align}
    \Lambda_x&=2t \, \big( P_x P_t+ e^{2P}Q_x Q_t\big),
    \\
    \label{eq:constraints2}
    \Lambda_t&=t \, \big( P_x^2+t e^{2P}Q_x^2 + P_t^2 + e^{2P}Q_t^2 \big).
  \end{align}
\end{subequations}

It turns out that
\Eqref{eq:evolLambda} can be ignored in the following sense. 
Given a time $t_0>0$, 
we can prescribe initial data $(P,Q)|_{t_0}$ for
the system \eqref{eq:originalgowdy} while assuming the condition
\begin{equation}
  \label{eq:initialdatacondition}
\int_0^{2\pi}(P_x P_t+e^{2P}Q_x Q_t) \, dx = 0 \qquad \text{ at } t=t_0.
\end{equation}
Then, the first constraint in \eqref{eq:constraints} determines the
function $\Lambda$ at the initial time, up to a constant which we
henceforth fix. Next, one easily checks that the solution $(P,Q)$ of
\eqref{eq:originalgowdy} corresponding to these initial data does
satisfy the compatibility condition associated with
\eqref{eq:constraints} and, hence, \eqref{eq:constraints} determines
$\Lambda$ uniquely for {\sl all} times of the evolution.  Moreover,
one checks that \eqref{eq:evolLambda} is satisfied identically by the
constructed solution $(P,Q,\Lambda)$. Consequently, equations
\eqref{eq:originalgowdy} represent the essential set of Einstein's
field equations for Gowdy spacetimes.
We refer to 
\eqref{eq:originalgowdy} as the \textit{Gowdy
  equations} and focus our attention on them.   
One could also consider the
alternative viewpoint which follows from the natural $3+1$-splitting
and treats the three equations
\eqref{eq:originalgowdy}-\eqref{eq:evolLambda} as an evolution system
for the unknowns $(P,Q,\Lambda)$, and \eqref{eq:constraints} will be regarded as constraints that propagate
if they hold on an initial hypersurface.

Rendall and collaborators \cite{AnderssonRendall, KichenassamyRendall,
  Rendall00}
   developed the so-called {\sl Fuchsian method} to
handle singular evolution equations derived from the Einstein
equations.  This method allows one
 to derive precise information about the behavior of solutions near the
singularity, which was a key step in the general proof of Penrose's
strong cosmic conjecture eventually established by Ringstr\"om
\cite{Ringstrom6}.  For the most recent developments we refer to \cite{ChoquetIsenberg,Choquet08,Choquet09}; 
especially, in \cite{Choquet08}, a generalization of the standard Fuchsian theory was recently introduced. 

In the first part \cite{BeyerLeFloch1}, we 
revisited the Fuchsian theory and developed another approach to the
 well-posedness theory in Sobolev spaces which applies to 
the singular initial value problem of a class of
(second-order, hyperbolic) Fuchsian systems. The theory in \cite{BeyerLeFloch1}
is particularly well-adapted to handle the Gowdy equations, as we show here. 
In fact, we propose here a rather simple proof of the 
well-posedness of the singular initial value problem for the Gowdy
equations, which provides an alternative to the approach in Rendall 
\cite{Rendall00}. Moreover, in passing, we are also able to clarify certain issues.

A second objective in the present paper is to present a new numerical approach and apply it specifically to Gowdy spacetimes. 
This approach is inspired by a pioneering work by Amorim, Bernardi, and LeFloch \cite{ABL}, which computed solutions to the Gowdy equations with data imposed on the singularity. 
The approximation scheme proposed in the present paper,  
which we refer to as the {\sl Fuchsian numerical algorithm,} 
is directly derived from our existence theory
 \cite{BeyerLeFloch1} and relies strongly 
on the hyperbolicity of the equations, so that error estimates for the numerical approximations are expected to hold.

This paper is organized as follows. \Sectionref{sec:theoreticalpart}
is devoted to the theoretical discussion of the singular initial value
problem for the Gowdy equations. First, we recall some heuristics for
the behavior of the solutions at the ``singularity'' $t=0$. Then, we
show that our notion of singular initial value problems in 
\cite{BeyerLeFloch1} is consistent with these heuristics, and then 
state the 
well-posedness results that follows from the first part. 
In \Sectionref{sec:numericalpart}, we introduce a general
numerical approach to solve second-order hyperbolic Fuchsian systems
numerically and then apply it to the Gowdy equations. Several numerical experiments are 
presented and, finally we numerically construct
 a Gowdy symmetric solution of Einstein's field
equations with a smooth non-compact Cauchy horizon.


\section{Singular initial value problem for the Gowdy equations}
\label{sec:theoreticalpart}

\subsection{Heuristics about the Gowdy equations}
\label{sec21} 

We provide here a formal discussion which motivates the following (rigorous) analysis.  
Based on extensive numerical experiments, it was first conjectured 
(and later established rigorously) that as one approaches the singularity
the spatial derivative of solutions $(P,Q)$ to \eqref{eq:originalgowdy} become negligible and $(P,Q)$,  
should approach a solution of the {\sl ordinary} differential equations 
\be
\aligned
& P_{tt} + {P_t \over t} = e^{2P} Q^2_t, 
\\
& Q_{tt} + {Q_t \over t} = - 2 \, P_t \, Q_t. 
\endaligned
\label{GS.ODE}
\ee
These equations are referred to, in the literature, as the {\sl velocity term dominated} (VTD) equations. 
Interestingly enough, they admit a large class of solutions given explicitly by 
\be
\label{PQ} 
\aligned 
& P(t,x) = \ln \big( \alpha \, t^{k} (1 + \zeta^2 t^{-2k})\big), 
\\
& Q(t,x) = \xi - {\zeta \, t^{-2k} \over \alpha \, (1 + \zeta^2 t^{-2k})},
\endaligned
\ee
where $x$ plays simply the role of a parameter and 
$\alpha, \zeta, \xi, k$ are arbitrary $2\pi$-periodic functions of $x$. 

Based on \eqref{PQ}, it is a simple matter to determine the first two terms in the expansion of the function $P$ near $t=0$, 
that is, for $k\not=0$ at least,
$$
\lim_{t \to 0} {P(t,x) \over \ln t} = \lim_{t \to 0} t \, P_t(t,x) = -|k|, 
$$
hence
$$
\lim_{t \to 0} \big( P(t,x) + |k (x)| \ln t \big) = \varphi(x), 
        \qquad
\varphi :=\begin{cases}
        \ln\alpha,     &  k < 0, 
        \\
        \ln (\alpha(1 + \zeta^2)),   & k = 0, 
        \\
        \ln (\alpha\zeta^2),   & k > 0.
        \end{cases}
$$
Similarly, for the function $Q$ we obtain 
$$
\aligned
& \lim_{t \to 0} Q(t,x) = q (x), 
        \qquad 
q := \begin{cases}
        \xi,   & k < 0, 
        \\
         \xi - \dfrac{\zeta}{\alpha(1+\zeta^2)},  &  k = 0,
        \\
         \xi - \dfrac{1}{\alpha \zeta}, & k > 0,
        \end{cases}
\\
& \lim_{t \to 0} t^{-2|k|} \, \big( Q(t,x) - q (x) \big) = \psi(x), 
        \qquad \psi :=\begin{cases}
        -\dfrac{\zeta}{\alpha}, &  k < 0, 
        \\
        \hskip.41cm 0, & k = 0,
        \\
        \hskip.18cm \dfrac{1}{\alpha\zeta^3}, & k >0.
        \end{cases}
\endaligned 
$$

From \eqref{PQ}, we thus have the expansion 
\be
\aligned
& P = -|k| \ln t + \varphi + o(1),
\\
& Q = q + t^{2|k|} \psi+o(t^{2|k|}), 
\endaligned
\label{GS.expansion}
\ee
in which $k, \varphi, q, \psi$ are functions of $x$. 
In general, $P$ {\sl blows-up to $+\infty$} when one approaches the singularity, while $Q$ 
remains {\sl bounded.}   
Observe that the sign of $k$ is irrelevant as far the asymptotic expansion is concerned, and we are allowed to restrict 
attention to $k\ge 0$.

By plugging the explicit solution in the nonlinear terms arising in \eqref{GS.ODE} one sees that 
$e^{2P} \, Q_t^2$ is of order $t^{2(|k|-1)}$ which is negligible since the left-hand side of the 
$P$-equation is of order $t^{-2}$, at least when $k \neq 0$. 
On the other hand, the nonlinear term $P_t Q_t$ is of order $t^{2(|k|-1)}$, which is the same order as the left-hand side
of the $Q$-equation. It is not negligible, but we observe that $P_t Q_t$ has the same behavior as $-(|k|/t) \, Q_t$. 

In fact, observe that the homogeneous system deduced from \eqref{GS.ODE}: 
\be
P_{tt} + {P_t \over t} = 0, \quad
Q_{tt} + {1 - 2 k \over t} Q_t = 0,
\label{GS.ODE-simple}
\ee
is solved precisely by the leading-order terms in \Eqref{GS.expansion}. 
This tells us that, as $t \to 0$, 
 the term $e^{2P} Q^2_t$ is negligible in the first equation in 
 \Eqref{GS.ODE}, 
 while $P_tQ_t + (|k|/t)Q_t$ is negligible at $t=0$.
This discussion hence allows us to conclude that 
as far as the behavior at the coordinate singularity $t=0$ is concerned, 
the nonlinear VTD equations \eqref{GS.ODE} are well approximated by the system \eqref{GS.ODE-simple}.

\ 

We return now to the nonlinear terms which were not included in the VTD equations, 
but yet are present in the full model \eqref{eq:originalgowdy}. 
 Allowing ourselves to differentiate the expansion \eqref{GS.expansion}, we
  get the following leading-order terms at $t=0$:
$$
e^{2P} \, Q_x^2 = \begin{cases}
        t^{-2|k|} e^{2\varphi}\, q_x^2 + \ldots,    & q_x \neq 0, 
        \\
        2e^{2\varphi}|k|_x \psi\ln t + \ldots,    & q_x=0,\quad |k|_x\not=0, 
        \\
        e^{2\varphi}\psi_x^2 + \ldots,      & q_x = 0,\quad |k|_x=0, \quad \psi_x \neq 0,  
        \end{cases} 
$$
$$
P_x \, Q_x 
=  \begin{cases}
-\ln t \, |k|_x \, q_x + \ldots,   &  |k|_x, q_x \neq 0, 
\\
\varphi_x \, q_x + \ldots,   &  |k|_x = 0, \quad \varphi_x, q_x \neq 0, 
\\
-2\ln^2 t\, t^{2|k|}\, (|k|_x)^2 \psi + \ldots,   &  |k|_x\neq 0,\quad q_x=0
\\
t^{2|k|} \varphi_x \, \psi_x + \ldots,   &  |k|_x= q_x = 0, \quad \varphi_x, \psi_x \neq 0. 
\end{cases} 
$$


To check (formally) the validity of the expansion \eqref{GS.expansion} we now return 
to the full system. Consider the nonlinear term $e^{2P} \, Q_x^2$ in \eqref{eq:originalgowdy}. 
and observe the following: 
\begin{itemize}
\item Case $q_x \neq 0$ everywhere on an open subinterval of $[0,2\pi]$. Then, on one hand,  
the left-hand side of the first equation in  \eqref{eq:originalgowdy}
 is of order $t^{-2}$, at most. 
On the other hand, the term $e^{2P} \, Q_x^2$ is negligible with respect to $t^{-2}$ 
if and only if the asymptotic velocity satisfies $|k| <1$ and is of the same order if $|k|=1$.  
\item Case $q_x = 0$ on an open subinterval of $[0,2\pi]$. Then,  
$e^{2P} \, Q_x^2$ is negligible with respect to $t^{-2}$, and no condition on the velocity 
$k$ is required on that interval. 
\item Case $q_x(x_0) = 0$ at some isolated point $x_0$. Then, no definite conclusion can be obtained 
and a ``competition'' between $|k|$ (which may approach the interval $[0,1]$) 
and $q_x$ (which approaches zero) is expected. 
\end{itemize} 

Similarly, at least when $|k|_x \, q_x \neq 0$, 
the nonlinear term $P_x Q_x$ is of order $\ln t$ and, therefore, 
negligible with respect to $t^{2(|k|-1)}$ (given by the left-hand side of the second equation in \eqref{eq:originalgowdy})
if and only if the asymptotic velocity is $|k| \leq 1$. Points where $|k|_x$ or $q_x$
vanish lead to a less singular behavior and condition on the velocity can also be relaxed.

The formal derivation above strongly suggests that we seek solutions to the full nonlinear equations 
admitting an asymptotic expansion of the form \eqref{GS.expansion}, that is 
$$
P = - k \, \ln t + \varphi + o(1),\quad
Q = q + t^{2 k} \, ( \psi + o(1)),  
$$
where $k \geq 0$ and $\varphi, q, \psi$ are prescribed. 
In other words, these solutions asymptotically approach a solution of the VTD equations and, in consequence, such solutions
will be referred to as {\sl asymptotically velocity term dominated} (AVTD) solutions.

Based on this analysis and extensive numerical experiments, it has 
been conjectured that asymptotically as one approaches the coordinate singularity $t=0$ the 
function $P(t,x)/\ln t$ should approach some limit $k=k(x)$, referred to as the 
\emph{asymptotic velocity,} and that $k(x)$ should belong to $[0,1)$ with the exception of 
a zero measure set of ``exceptional values''.


\subsection{Gowdy equations as a second-order hyperbolic Fuchsian system}

The first step in our (rigorous) analysis of the Gowdy equations
\Eqsref{eq:originalgowdy} now is to write them as a system of
second-order hyperbolic Fuchsian equations. These
are equations of the form (written here for a scalar field in order to keep the notation simple)
\begin{equation}
  \label{eq:2ndorderhypFuchs}
  D^2 v(t,x)+2a(x) Dv(t,x)+b(x) v(t,x)
  =t^2 c^2(t,x)\del_x^2 v(t,x)+f(t,x,v,Dv,\del_x v),
\end{equation} 
where $v:(0,\delta]\times U\to\R$ is the unknown defined on an interval 
$U\subset\R$ an interval (with $\delta>0$). Here we use the
symbol $D:=t\del_t$, which denotes the time
derivative operator and is singular at the origin $t=0$; by $D^2 v$ we
mean $D(Dv)=t\del_t(t\del_t v)$. We assume that all quantities
are periodic in $x$ with a periodicity domain $U$, and for the present applications we set $U:=[0,2\pi]$. 
The coefficients $a$ and $b$ are smooth and
depend on the spatial coordinate $x$, only. The characteristic speed
$c$ has to satisfy certain properties at $t=0$; however, since in our
application we have $c\equiv 1$, we do not need to discuss those
here. All further details are explained in \cite{BeyerLeFloch1}. 
The left-hand side of the equation is called the \textit{principal part},
and its
right-hand side the \textit{source-term}. In the course of the discussion, we
will often abuse notation slightly and refer to the function $f$
alone as the source-term.

In the first part \cite{BeyerLeFloch1}, we determined 
leading-order behavior of solutions to \eqref{eq:2ndorderhypFuchs} at $t=0$, 
from
the assumptions that it is ``driven'' by the principal part of the equation
(in a well-defined manner) 
and, additionally, the source-term has a suitable decay property at
$t=0$. In particular, spatial derivatives are
negligible, in a sense made precise in \cite{BeyerLeFloch1}. 

From the equation \eqref{eq:2ndorderhypFuchs}, we define 
$$
  u_0(t,x):=\begin{cases}
    u_*(x)\,t^{-a(x)}\ln t+u_{**}(x)\,t^{-a(x)}
    & \quad a^2=b,\\
    u_*(x)\,t^{-\lambda_1(x)}+u_{**}(x)\,t^{-\lambda_2(x)},
    & \quad a^2 \neq b,
  \end{cases}
$$
and
\begin{equation}
  \label{eq:deflambda2}
  \lambda_{1}:=a+\sqrt{a^2-b},\quad \lambda_{2}:=a-\sqrt{a^2-b}.
\end{equation}
Now, $u_0$ is regarded a prescribed (real-valued) function 
and, as far as our application to \eqref{eq:originalgowdy} is concerned,
$\lambda_1$ and $\lambda_2$ are real-valued. The function $u_0$ is smooth on $(0,\delta]\times U$, provided
either 
$a^2(x)\not=b(x)$ for all $x\in U$ or else 
$a^2(x)=b(x)$ for all $x\in
U$. More generally, when there is a transition between these two regimes,
the functions $u_*$ and $u_{**}$ have to be renormalized
\cite{BeyerLeFloch1} in order to guarantee the smoothness of $u_0$, which we assume from now on. The
function $u_0$ represents the leading-order terms in the
``\textit{singular initial value problem with two-term asymptotic data
  $u_*$ and $u_{**}$}'', as discussed in \cite{BeyerLeFloch1} and is
referred to as a \textit{canonical two-term expansion}.

After multiplication by $t^2$, the equations \eqref{eq:originalgowdy} immediately
take the second-order hyperbolic Fuchsian form
\begin{align*}
  D^2P&=t^2\del_x^2 P+e^{2P} (DQ)^2 - t^2 e^{2P}(\del_x Q)^2,
  \\
D^2Q&=t^2\del_x^2 Q-2 DP DQ + 2t^2 \del_x P \del_x Q.
\end{align*} 
The general canonical two-term expansion then reads  
$$
P(t,x)=P_*(x)\log t+P_{**}(x)+\ldots 
$$
for the function $P$ 
and, similarly, an expansion $Q_*(x)\log t+Q_{**}(x)+\ldots$ for the function $Q$ with prescribed data $Q_*, Q_{**}$.
At this stage, we do not make precise statements
about the (higher-order) remainders, yet. In any case, the
theory in \cite{BeyerLeFloch1} does not apply to this system directly, 
due to the presence of the term $-2 DP DQ$ ---with the exception of the cases $P_*=0$ or $Q_*=0$. 
Namely, this term
does not behave as a positive power of $t$ at $t=0$ when we
substitute $P$ and $Q$ by their canonical two-term expansions, but this 
is required by the theory. 

At this juncture, motivated by the formal discussion in Section~\ref{sec21}, especially \eqref{GS.ODE-simple}, 
we propose to 
add a term $-2k DQ$ to the equation for $Q$ where $k$ is a prescribed (smooth, 
spatially periodic) function depending on $x$, only. This yields the system of equations 
\begin{equation}
  \label{eq:Fuchsiangowdy}
  \begin{split}
    D^2P&=t^2\del_x^2 P+e^{2P} (DQ)^2 - t^2 e^{2P}(\del_x Q)^2,\\
    D^2Q-2k DQ&=t^2\del_x^2 Q-2 (k+DP) DQ 
    + 2t^2 \del_x P \del_x Q.
  \end{split}
\end{equation}
Later, the function $k$ will play the role of the asymptotic velocity
mentioned before. The resulting system is of second-order hyperbolic
Fuchsian form with two equations, corresponding to 
\be
\label{809}
\lambda_1^{(1)}=\lambda_2^{(1)}=0,
\quad\lambda_1^{(2)}=0,\,\lambda_2^{(2)}=-2k.
\ee
Here, the superscript determines the respective equation of the system
\Eqsref{eq:Fuchsiangowdy}.  If we assume that $k$ is a strictly
positive function, as we will do in all what follows, the expected
leading-order behavior at $t=0$ given by the canonical two-term
expansions is
\begin{equation}
  \label{eq:leadingorderPQ}
  \aligned
  P(t,x)& =P_*(x)\log t+P_{**}(x)+\ldots,
  \\
  Q(t,x)& =Q_*(x)+Q_{**}(x)t^{2k(x)}+\ldots 
\endaligned
\end{equation}
 One checks easily that the problem, which we had for the
previous form of the equations, does not arise if $P_*=-k$. Indeed,
the canonical two-term expansion is consistent with the heuristics of
the Gowdy equations above and we recover the singular initial value
problem studied rigorously in \cite{KichenassamyRendall,Rendall00} and
numerically in \cite{ABL}. We only mention here without further notice
that the case $k=0$ with the logarithmic canonical two-term expansion
for $Q$ given by the first form of the equations above is covered by
the following discussion. Furthermore, 
the case of possibly vanishing $k$ may be also included via a suitable 
normalization of the asymptotic data \cite{BeyerLeFloch1}.


\section{Well-posedness theory for the Gowdy equations}
\label{sec:wellposednessGowdy}

\subsection{Reformulation of the problem}
\label{311} 

When $P_*=-k$, this function plays a two-fold role in
\Eqref{eq:Fuchsiangowdy}. On one hand, it is an
asymptotic data for the function $P$ and, on the other hand, it is 
a coefficient of the equation satisfied by the function $Q$. In order to keep
these two roles of $k$ separated in a first stage and instead
of \Eqref{eq:Fuchsiangowdy}, we consider the system
\begin{equation}  
  \label{eq:FuchsiangowdyManip}
  \begin{split}
    D^2P&=t^2\del_x^2 P+e^{2P} (DQ)^2 - t^2 e^{2P}(\del_x Q)^2,\\
    D^2Q-2k DQ&=t^2\del_x^2 Q-2 (-P_*+DP) DQ 
    + 2t^2 \del_x P \del_x Q.
  \end{split}
\end{equation}
Studying the singular initial value problem with two-term asymptotic
data means that we search for solutions to \Eqref{eq:FuchsiangowdyManip}
of the form (as $t \to 0$) 
\begin{equation}
  \label{eq:expansionGowdy}
  \begin{split}
    P(t,x)&=P_*(x)\log t+P_{**}(x)+w^{(1)}(t,x), \\
    Q(t,x)&=Q_*(x)+Q_{**}(x)t^{2k(x)}+w^{(2)}(t,x),
  \end{split}
\end{equation}
for general asymptotic data $P_*$, $P_{**}$, $Q_*$, $Q_{**}$,
and remainders $w^{(1)}$, $w^{(2)}$ which will be specified to be suitably ``small''
and belong to
 certain functional spaces; see
below. After studying the well-posedness for this problem, we can always choose $P_*$
to coincide with $-k$ 
and, therefore, recover our original Gowdy problem
\eqref{eq:Fuchsiangowdy}-\eqref{eq:leadingorderPQ}. For simplicity in the presentation, we always assume 
that $k$ is a $C^\infty$ function.

In the following discussion, we write the vector-valued remainder as
$w:=(w^{(1)},w^{(2)})$, and we fix some asymptotic data $P_*$, $P_{**}$, $Q_*$, and
$Q_{**}$. In agreement with the notation in \cite{BeyerLeFloch1}, the source-term operator $F=(F_1,F_2)$ is defined by
$$
F[w](t,x):=f\Big(t,x,u_0+w,D(u_0+w),\del_x(u_0+w)\Big),
$$ 
where $u_0$ is the vector-valued, canonical two-term expansion given by
\Eqref{eq:expansionGowdy} and the asymptotic data. We write
$$
F[w](t,x)=:(F_1[w](t,x),F_2[w](t,x)),
$$ 
and from \Eqsref{eq:FuchsiangowdyManip} we obtain 
\begin{align*}
  F_1[w]=&\left(t^{P_*} e^{P_{**}} e^{w^{(1)}} \big( 2k\, t^{2k}Q_{**}+Dw^{(2)} \big) \right)^2
  \\
  &-\left(t^{P_*} e^{P_{**}} e^{w^{(1)}} \, \big( t \, \del_x Q_*
    +2\del_x k t^{2k}t\ln t Q_{**}
    + t^{2k}t\del_xQ_{**}+t\del_xw^{(2)} \big) \right)^2
\end{align*}
and
$$
\aligned
F_2&[w]
\\
= &-2 Dw^{(1)} \, \big( 2k\, t^{2k}Q_{**}+Dw^{(2)} \big)
\\
 & +2 \Big( t \del_xP_*\log t+t\del_x P_{**}+t\del_xw^{(1)} \Big)
     \Big( \big(t \del_x Q_* + 2 \, \del_x k t^{2k}t\ln t Q_{**}
  + t^{2k}t\del_xQ_{**}+t\del_xw^{(2)} \Big).
\endaligned
$$


\subsection{Properties of the source-term operator}

To establish the well-posedness of the singular initial value
problem for the Gowdy equations,
 we need first to derive certain decay properties of the source-term
operator $F$.  In \cite{BeyerLeFloch1}, we introduced the spaces $X_{\delta,\alpha,k}$ and
$\Xt_{\delta,\alpha,k}$ associated with constants $\delta,\alpha>0$ and
non-negative integers $k$: in short, a function $w$ belongs to
 $X_{\delta,\alpha,k}$ if each derivative $\del_x^l D^m
w$ (with $l+m\le k$) weighted by 
$t^{\lambda_2(x)-\alpha}$ is a bounded continuous map
$(0,\delta]\to L^2(U)$. Here, $\lambda_2$ given by \Eqref{eq:deflambda2}
is determined by the
coefficients of the second-order hyperbolic Fuchsian equation under consideration. 
The corresponding norm
$\|\cdot\|_{\delta,\alpha,k}$ turns
$(X_{\delta,\alpha,k},\|\cdot\|_{\delta,\alpha,k})$ into a Banach
space. For any $w\in X_{\delta,\alpha,k}$, this norm has the form 
$\|w\|_{\delta,\alpha,k}=\sup_{t\in(0,\delta]}
E_{\delta,\alpha,k}[w](t)$, 
where $E_{\delta,\alpha,k}[w]:(0,\delta]\to \R$  is
bounded and continuous.  

The spaces $\Xt_{\delta,\alpha,k}$ and the associated maps $\Et_{\delta,\alpha,k}$ are defined analogously;
only the weight of the $k$-th spatial derivative is substituted by\footnote{The 
(slightly) more general definition given in \cite{BeyerLeFloch1} 
involved a speed coefficient $c$, taken here to be identically unit.} 
$t^{\lambda_2(x)-\alpha+1}$. 
It always follows that $X_{\delta,\alpha,k}\subset\Xt_{\delta,\alpha,k}$.
There is, in fact, no reason to assume $\alpha>0$ to remain constant 
in the definition of these spaces, since no essentially new
difficulty arises in the theory \cite{BeyerLeFloch1} when 
$\alpha$ is a (spatially periodic) strictly positive function in
$C^1(U)$.

Let us introduce some further
notation specific to the Gowdy equations. 
Let $X_{\delta,\alpha_1,k}^{(1)}$ be the space defined as above and based 
on the coefficients of the first equation in \Eqsref{eq:FuchsiangowdyManip}
and, similarly, let $X_{\delta,\alpha_2,k}^{(2)}$ be the space associated with 
the second equation. By definition, 
a vector-valued map $w= (w^{(1)}, w^{(2)})$ belongs $X_{\delta,\alpha,k}$
precisely if $w^{(1)}\in X_{\delta,\alpha_1,k}^{(1)}$ and $w^{(2)}\in
X_{\delta,\alpha_2,k}^{(2)}$, with $\alpha:=(\alpha_1,\alpha_2)$. 
An analogous notation is used for the spaces $\Xt_{\delta,\alpha_1,k}^{(1)}$, $\Xt_{\delta,\alpha_2,k}^{(2)}$ and
$\Xt_{\delta,\alpha,k}$.

Now we are ready to state a first result about the source-term of \Eqsref{eq:FuchsiangowdyManip}. 

\begin{lemma}[Operator $F$ in the finite differentiability class]
  \label{lem:finitediffF}
Fix any $\delta>0$ and any asymptotic data 
$$
P_*, P_{**},Q_*,Q_{**}\in H^m(U) \qquad m\ge 2. 
$$ 
Suppose there exist $\eps>0$ and a continuous function
  $\alpha=(\alpha_1,\alpha_2) : U \to (0,\infty)^2$ so that, at each $x\in U$, 
  \begin{subequations}
    \label{eq:finitediffF}
    \begin{gather}
      \label{eq:finitediffF1}
      \alpha_1(x)+\eps<
      \min\big( 2(P_*(x)+2k(x)),2(P_*(x)+1)\big),
      \\
      \label{eq:finitediffF2}
      \alpha_2(x)+\eps<2(1-k(x)),\\
      \label{eq:finitediffF3}
      \alpha_1(x)-\alpha_2(x)>\eps+\min\big( 0,2k(x)-1\big),
      \\
      \eps<1.
    \end{gather}
  \end{subequations}
  Then, the operator $F$ associated with the system \eqref{eq:FuchsiangowdyManip} and the 
  given asymptotic data 
  maps 
  $\Xt_{\delta,\alpha,m}$ into $X_{\delta,\alpha+\eps,m-1}$ 
and satisfies the
  following Lipschitz continuity condition: For each $r>0$ and for some 
  constant $C>0$ (independent of $\delta$), 
$$
E_{\delta,\alpha+\eps,m-1}\Big[ F[w]-F[\what] \Big](t)
  \le C \, \Et_{\delta,\alpha,m}[w-\what](t), \qquad t\in (0,\delta]
$$
  for all $w,\what\in B_r\subset
  \Xt_{\delta,\alphat,m}$, where $B_r$ denotes the closed ball centered at the origin. 
\end{lemma}

In this lemma, since $P_*\in H^1(U)$, in particular, a standard Sobolev
inequality implies that $P_*$ can be identified with a unique bounded
continuous periodic function on $U$, and the inequality \eqref{eq:finitediffF1} makes sense pointwise.

\begin{proof} Consider the expression of $F$ given at the end of Section~\ref{311}. 
  Let $w\in\Xt_{\delta,\alpha,m}$ for some (so far unspecified)
  positive spatially dependent functions $\alpha_1,\alpha_2$, hence
  $w^{(1)}\in\Xt_{\delta,\alpha_1,m}^{(1)}$ and $w^{(2)}\in\Xt_{\delta,\alpha_2,m}^{(2)}$. 
  By a standard Sobolev inequality (since $m \geq 2$ and the 
  spatial dimension is $1$), 
  we get that $F[w](t,\cdot)\in H^{m-1}(U)$ for all
  $t\in(0,\delta]$.  Namely, if $m\ge 2$ we can control the non-linear
  terms of $F[w](t,\cdot)$ in all generality for a given $t>0$ if any
  factor in any term of $F[w](t,\cdot)$, after applying up to $m-1$
  spatial derivatives, is an element in $L^\infty(U)$ -- with the
  exception of the $m$th spatial derivative of $w$ which is only
  required to be in $L^2(U)$. This is guaranteed by the Sobolev
  inequalities.  Having found that $F[w](t,\cdot)\in H^{m-1}(U)$ for
  all $t\in(0,\delta]$, it is easy to check that $F_1[w]\in
  X_{\delta,\alpha_1+\eps,0}^{(1)}$ if
  \begin{equation}
    \label{eq:firstconditionPre}
    \alpha_1(x)+\eps\le
    \min\Big( 2(P_*(x)+2k(x)),2(P_*(x)+1)\Big), \qquad  x\in U.
  \end{equation}
  Even more, condition \Eqref{eq:firstconditionPre} implies that
  $D^lF_1[w]\in X_{\delta,\alpha_1+\eps,0}^{(1)}$ for all $l\le
  m-1$. 
  
  Considering now spatial derivatives, we have to deal
  with two difficulties. The first one is that  
  logarithmic terms arise with each spatial derivative. We find 
  $\del_x^kD^lF_1[w]\in X_{\delta,\alpha_1+\eps,0}^{(1)}$ for
  all $l\le m-1$ and $k\le m-2$ and $k+l\le m-1$ (excluding first the case $k=m-1$, $l=0$)
   provided
  \begin{equation}
    \label{eq:firstcondition}
    \alpha_1(x)+\eps<
    \min\Big( 2(P_*(x)+2k(x)),2(P_*(x)+1)\Big),\qquad x\in U.
  \end{equation}

A second difficulty arises in the case $k=m-1$, $l=0$. Namely, since 
  $w\in\Xt_{\delta,\alpha,m}$ (and not in $X_{\delta,\alpha,m}$),
  it follows that in particular $t\del^m_x w^{(2)}\sim
  t^{2k+\alpha_2}$ (and not $t^{1+2k+\alpha_2}$); note that the
  function $\beta$ which determines the behavior of the characteristic
  speeds at $t=0$ in \cite{BeyerLeFloch1} is identically zero in the
  case of the Gowdy equations. The potentially problematic term is
  hence of the form $AB$ with 
  \begin{align*}
    A:= &t^{P_*} e^{P_{**}} e^{w^{(1)}}(t\del_x Q_*
    +2\del_x k t^{2k}t\ln t Q_{**}
    + t^{2k}t\del_xQ_{**}+t\del_xw^{(2)}),
    \\
    B:=&\cdot t^{P_*} e^{P_{**}} e^{w^{(1)}}\bigl(\del_x^{m-1}(t\del_x Q_*
    +2\del_x k t^{2k}t\ln t Q_{**}
    + t^{2k}t\del_xQ_{**})+t\del_x^{m}w^{(2)}\bigr),
  \end{align*}
  originating from taking $m-1$ spatial derivatives of $F_1[w]$. 
  To ensure $\del_x^{m-1}F_1[w]\in
  X_{\delta,\alpha_1+\eps,0}^{(1)}$, we need 
  \begin{equation}
    \label{eq:secondcondition}
    \alpha_1(x)+\eps< (P_*(x)+1)+(P_*(x)+2k(x)+\alpha_2(x)),
    \qquad x\in U.
  \end{equation}
  If \Eqref{eq:firstcondition} is satisfied, we have (for all $x$) 
$$
  \aligned
    \alpha_1(x)+\eps
    & <
    \min\Big( 2(P_*(x)+2k(x)),2(P_*(x)+1)\Big) 
 \\
    & \le (P_*(x)+1)+(P_*(x)+2k(x))
\endaligned
$$
  and, thus, \Eqref{eq:secondcondition} follows from
  \Eqref{eq:firstcondition}. In conclusion, 
  \Eqref{eq:firstcondition} is sufficient to guarantee that $F_1[w]\in
  X_{\delta,\alpha_1+\eps,m-1}^{(1)}$. 
  
  Let us proceed next with the analysis of the term 
  $F_2[w]$. If
  \begin{equation}
    \label{eq:thirdcondition}
    \alpha_1(x)-\alpha_2(x)\ge\eps,
    \quad\qquad 
    \alpha_2(x)+\eps<2(1-k(x)),\qquad x\in U,
  \end{equation}
  then $F_2[w]\in X_{\delta,\alpha_2+\eps,0}^{(2)}$. This
  inequality also implies that all time derivatives are in
  $X_{\delta,\alpha_2+\eps,0}^{(2)}$ as before. We have to deal
  with the same two difficulties as before when we consider spatial
  derivatives of $F_2[w]$. On one hand, 
  equality in
  \Eqref{eq:thirdcondition} cannot occur due to additional
  logarithmic terms. On the other hand, we must be careful with the $(m-1)$-th
  spatial derivative of $F_2[w]$.  Here, the two problematic terms 
  are of the form $AB$ with either 
  \begin{align*}
   A:= &\del_x^{m-1}(t\del_xP_*\log t+t\del_x P_{**})
    +t\del_x^m w^{(1)},
    \\
    B:= &t\del_x Q_*
    +2\del_x k t^{2k}t\ln t Q_{**}
    + t^{2k}t\del_xQ_{**}+t\del_xw^{(2)},
  \end{align*}
or else 
  \begin{align*}
    A:= & t\del_xP_*\log t+t\del_x P_{**}+t\del_x w^{(1)},
    \\
    B:= &\del_x^{m-1}(t\del_x Q_*
    +2\del_x k t^{2k}t\ln t Q_{**}
    + t^{2k}t\del_xQ_{**})+t\del_x^mw^{(2)}.
  \end{align*}
  The first one is under control provided
  $\alpha_1(x)+1>2k(x)+\alpha_2(x)+\eps$, for all $x\in U$, while
  for the second one it is sufficient to require $\eps<1$.  The
  claimed Lipschitz continuity condition follows from the above arguments.
\end{proof}

Obviously, positive functions $\alpha_1$ and $\alpha_2$ and constants
$\eps>0$ satisfying the hypothesis of \Lemref{lem:finitediffF} can
exist only if $k(x)<1$ for all $x\in U$ (due to
\Eqref{eq:finitediffF2}) except for special choices of data; cf.~Lemma\ref{444}, below. 
Hence, we make the assumption that $0<k(x)<1$
for all $x$, which is consistent with our formal analysis in
 \Sectionref{sec21}.  As a consistency
check for the case of interest $P_*=-k$, let us determine under which
conditions the inequalities \Eqsref{eq:finitediffF} can be hoped to be
satisfied at all. For this, consider \Eqsref{eq:finitediffF1} and
\eqref{eq:finitediffF3} in the ``extreme'' case
$\alpha_2=\eps=0$. This leads to the condition $0<k<3/4$, which
shows that \Lemref{lem:finitediffF} does not apply within the full
interval $0<k<1$. 

It is interesting to note that Rendall was led to
the same restriction in \cite{Rendall00}, but its origin was not obvious
in his approach. Here, we find that this is caused by the presence of
the condition \eqref{eq:finitediffF3} in particular which reflects the
fact that $w$ is an element of the space $\Xt_{\delta,\alpha,m}$ rather than of the smaller space 
$X_{\delta,\alpha,m}$. Interestingly, we can eliminate this 
condition and, hence, retain the full interval $0<k<1$, when we consider
the $C^\infty$-case, instead of finite differentiability, as we now show.

\begin{lemma}[Operator $F$ in the $C^\infty$ class. General theory]
  \label{lem:CinftyF}
Fix any $\delta>0$ and any asymptotic data 
  \[P_*, P_{**},Q_*,Q_{**}\in C^\infty(U).
  \] 
  Suppose there exist a constant $\eps>0$ and a continuous
  functions $\alpha=(\alpha_1,\alpha_2) :U\to (0,\infty)^2$ such  that, 
  at each $x\in U$, 
  \begin{subequations}
    \begin{gather}
      \label{eq:CinftyF1}
      \alpha_1(x)+\eps<
      \min\big( 2(P_*(x)+2k(x)),2(P_*(x)+1)\big),
      \\
      \label{eq:CinftyF2}
      \alpha_2(x)+\eps<2(1-k(x)),\\
      \label{eq:CinftyF3}
      \alpha_1(x)-\alpha_2(x)>\eps.
    \end{gather}
  \end{subequations}
  Then, for each integer $m \ge 1$, the operator $F$ maps 
  $X_{\delta,\alpha,m}$ into $X_{\delta,\alpha+\eps,m-1}$
   and satisfies the
  following Lipschitz continuity property:
  for each $r>0$ and some constant $C>0$ (independent of $\delta$), 
$$
  E_{\delta,\alpha+\eps,m-1}\big[ F[w] - F[\what]\big](t)
  \leq
   C \Et_{\delta,\alpha,m}[w-\what](t), \qquad t\in (0,\delta], 
$$ 
  for all $w,\what\in B_r \cap
  X_{\delta,\alpha + \eps,m}\subset\Xt_{\delta,\alpha+ \eps,m}$. 
\end{lemma}

The proof is completely analogous to that of
\Lemref{lem:finitediffF}. Recall from \cite{BeyerLeFloch1} that only
spaces without the tilde are necessary for the well-posedness theory
in the $C^\infty$-case and hence that we obtain stronger control than
in the finite differentiability case.  Hence, the $C^\infty$-case does
not require the condition \eqref{eq:finitediffF3}. This has the
consequence that $k$ can have values in the whole interval $(0,1)$ as
we show in detail later.  In a special case, which will be of interest
for the later discussion, however, we can relax the constraints for
$k$ even in the finite differentiability case.

\begin{lemma}[Operator $F$ in the finite differentiability class. A special case]
\label{444} 
Fix any $\delta>0$ and any asymptotic data 
$$
P_*, P_{**},Q_{**}\in H^m(U), \quad Q_*=const, \qquad m \geq 2. 
$$ 
Suppose there exist 
  $\eps>0$ and 	a continuous function
  $\alpha=(\alpha_1,\alpha_2) : U \to (0,\infty)^2$ such that, at each $x\in U$, 
  \begin{align*}
    &\alpha_1(x)+\eps<2(P_*(x)+2k(x)),
    \\
    &\alpha_2(x)+\eps<2,
    \\
    &\alpha_1(x)-\alpha_2(x)>\eps-1, 
   \\
    &\eps<1.
  \end{align*} 
  Then, the operator $F$ satisfies the conclusions of \Lemref{lem:finitediffF}.
\end{lemma}

In the special case of constant asymptotic data
$Q_*=const$, we can prove the required properties of $F$ if the
function $k$ is any positive function in the finite differentiability
case.  The analogous result for the $C^\infty$-case can also be derived.


\subsection{Well-posedness of the singular initial value problem}

Relying on Theorem~3.10 in \cite{BeyerLeFloch1}, 
we now determine conditions that ensure that the singular initial value problem
for the Gowdy equations is well-posed.  Besides the properties 
of the source-operator $F$ already discussed, we
have to check the positivity of the \textit{energy dissipation matrix}
$$ 
  N:=
  \begin{pmatrix}
    \Re(\lambda_1-\lambda_2)+\alpha & ((\Im\lambda_1)^2/\eta-\eta)/2 & 0 \\
    ((\Im\lambda_1)^2/\eta-\eta)/2 
    & \alpha 
    & t\del_x{c}-\del_x\Re(\lambda_1-\lambda_2)(t c\ln t) \\
    0 & t\del_x{c}-\del_x\Re(\lambda_1-\lambda_2)(t c\ln t) 
    & \Re(\lambda_1-\lambda_2)+\alpha-1-Dc/c
  \end{pmatrix} 
$$
for some well-chosen constant $\eta>0$ 
and 
for each of the two Gowdy equations. Here, we omit the upper indices
order to simplify the notation,
while $\Re$ and $\Im$ denote the real and imaginary part
 of a complex number. In the present paper, the 
characteristic speed $c$ in \cite{BeyerLeFloch1} is constant equal to $1$, and all eigenvalues are real, so that the above matrix simplifies: 
$$ 
  \begin{pmatrix}
    \lambda_1-\lambda_2+\alpha & -\eta/2 & 0 \\
    -\eta/2 
    & \alpha 
    & -\del_x (\lambda_1-\lambda_2)(t \ln t) \\
    0 & -\del_x (\lambda_1-\lambda_2)(t \ln t) 
    & \lambda_1-\lambda_2+\alpha-1
  \end{pmatrix} 
$$
In view of \eqref{809}, this leads us to the matrix 
\be
\label{eq:defNfinal3-1}
  N^{(1)} :=
  \begin{pmatrix}
    \alpha_1 & -\eta/2 & 0 \\
    -\eta/2 
    & \alpha_1 
    & 0 \\
    0 & 0 
    &  \alpha_1-1 
  \end{pmatrix} 
\ee
for the first component and to the matrix 
\be
\label{eq:defNfinal3-2}
  N^{(2)}  :=
  \begin{pmatrix}
    2 k + \alpha_2 & -\eta/2 & 0 \\
    -\eta/2 
    & \alpha_2 
    & - 2 \del_x k (t \ln t) \\
    0 & - 2 \del_x k (t \ln t) 
    & 2 k +\alpha_2-1 
  \end{pmatrix} 
\ee
for the second component.

For the matrix $ N^{(1)}$ to be positive, it is necessary that
$\alpha_1(x)>1$ for all $x\in U$. However, if $P_*=-k$, then condition
\eqref{eq:finitediffF1} in \Lemref{lem:finitediffF} in the finite
differentiability case 
(or the corresponding one in
\Lemref{lem:CinftyF} in the $C^\infty$-case) implies that 
 $\alpha_1(x)<1$. Hence, in the same way
as  in Rendall \cite{Rendall00}, one does not arrive at a well-posedness result for
the singular initial value problem yet. 
However, since the
positivity of the energy dissipation matrix is the only part of the
hypothesis in \Theoremref{P1-th:well-posednessSIVP} of
\cite{BeyerLeFloch1} which is is violated, we can use instead
\Theoremref{P1-th:WellPosednessHigherOrderSIVP} of
\cite{BeyerLeFloch1} to prove well-posedness of the ``\textit{singular initial
value problem with asymptotic solutions of sufficiently high order}''.


Let us quickly recapitulate the basics for this singular initial value
problem which are discussed in detail in the first Part of this series. 
Consider any 
second-order  hyperbolic Fuchsian equation of the form
\eqref{eq:2ndorderhypFuchs} and, for any given asymptotic data, 
define
$$
\widehat F[w]:=F[w]+t^2 k^2\del_x^2 (u_0+w). 
$$ 
Let $H$ be the operator which maps any source function $f_0=f_0(t,x)$
(having a suitable behavior at $t=0$) to the remainder $w=v-u_0$, where
$v$ the unique solution of the ordinary differential equation
$$
D^2 v(t,x)+2a(x) Dv(t,x)+b(x) v(t,x)
=f_0(t,x),
$$ 
consistent with the prescribed asymptotic data. Finally, set $G:=H\circ\hat
F$. As is easily checked, $w$ is the remainder of a solution of the
full equations (consistent with the prescribed asymptotic data) if and only if
$w=G[w]$, that is, if $w$ is a fixed point of the map $G$. 

Set $w_1=0$ and define the sequence 
$$
w_{j+1}=G[w_j],\qquad j= 1,2, \ldots
$$ 
The convergence of this sequence to a fixed point is known for analytic data and 
for ordinary differential equations, only. Yet, the sequence $(w_j)$ has certain useful
properties. 

On one hand, the residual of the second-order
hyperbolic Fuchsian equation, i.e.~the difference between the left- and the
right-hand side is of higher order in $t$ (at $t=0$) if $j$
is larger. Hence, the sequence satisfies the original equations at a higher order of approximation 
if we choose larger values of $j$. A disadvantage is that the higher
we choose $j$, the more spatial derivatives of the asymptotic data
we need to control.  In any case, the main point for the current discussion of well-posedness for
 the Gowdy equations is the
following one.

Define the exponent 
\begin{equation}
  \label{eq:increasealpha}
  \alphat:=\alpha+(j-2)\kappa\eps,
\end{equation}
where $\alpha$ and $\eps$ are the quantities introduced above and
$\kappa<1$ is a constant which we can choose arbitrarily. If $v=(P,Q)$
is a solution of the Gowdy equations corresponding to given asymptotic
data, we set 
$$
w:=v-u_0-w_j
$$
for some $j\ge 1$. Then it follows from
our considerations in the first paper that the equation has a unique
solution with remainder $w\in X_{\delta, \alphat,k}$ for some
$k$, provided $j$ is large enough so that the energy dissipation
matrix (evaluated with $\alphat$) is positive.  Hence, our
previous discussion implies that the singular initial value problem
with asymptotic solutions of order $j$ is well-posed provided one of
the previous lemmas applies. (This is only true if the
asymptotic data functions are sufficiently regular.) 

We can be more specific about what we mean by $j$ being
``sufficiently large'', and we now make some choice for the parameters 
$\alpha_1$, $\alpha_2$ and $\eps$, consistent with
\Lemref{lem:CinftyF}, which will allow us to estimate the required
size of $j$. We make no particular effort to choose these
quantities optimally, but still the goal is to choose $j$ ``reasonably'' 
small. Henceforth, we restrict to the $C^\infty$-case and 
$P_*(x)=-k(x)$ with $0<k(x)<1$ for all $x\in U$. 
We introduce
positive constants $\mu_1$ and $\mu_2$ (with further restrictions
later) and the function $\chi(x):=1-2|x-1/2|$. The condition
\eqref{eq:CinftyF1} states that we must choose $\alpha_1(x)$ and
$\eps$ so that $\alpha_1(x)+\eps<\chi(k(x))$. We set
\begin{equation}
  \label{eq:choicealpha1}
  \alpha_1(x):=1-\sqrt{4(k(x)-1/2)^2+\mu_1^2},
\end{equation}
and find $\chi(k(x))-\alpha_1(x)>\sqrt{1+\mu_1^2}-1$ for all $x\in
U$, provided $0<k(x)<1$. Similarly, we set
\begin{equation}
  \label{eq:choicealpha2}
  \alpha_2(x):=1-\sqrt{4(k(x)-1/2)^2+\mu_2^2},
\end{equation}
and it follows that
$\alpha_1(x)-\alpha_2(x)>\sqrt{1+\mu_2^2}-\sqrt{1+\mu_1^2}$ for
$\mu_2>\mu_1$. For the conditions \eqref{eq:CinftyF1}
and \eqref{eq:CinftyF3} to hold true, we have to choose
\[0<\mu_1<\mu_2,\quad\text{and}\quad
0<\eps\le\min\Biggl(
\sqrt{1+\mu_1^2}-1,\sqrt{1+\mu_2^2}-\sqrt{1+\mu_1^2}
\Biggr).
\] 
Condition \eqref{eq:CinftyF2} is then satisfied automatically. 

Now, assume in what follows that $k(x)\in (1/2-\Delta
k,1/2+\Delta k)$ for all $x\in U$ for a constant $\Delta k\in
(0,1/2)$. Then it is clear that both functions $\alpha_1$ and
$\alpha_2$ are positive for all such $k(x)$ if and only if
\[\mu_1<\mu_2<\sqrt{1-4(\Delta k)^2}.\]
This assumption will be made in the following. In
\Theoremref{P1-th:WellPosednessHigherOrderSIVP} of
\cite{BeyerLeFloch1}, we could choose $j$ as small as
possible if we pick the maximal allowed value for $\eps$. Hence,
we set
\[\eps:=\min\Biggl(\sqrt{1+\mu_1^2}-1,\sqrt{1+\mu_2^2}-\sqrt{1+\mu_1^2}
\Biggr).
\]
We find easily that
\[\sqrt{1+\mu_1^2}-1\le \sqrt{1+\mu_2^2}-\sqrt{1+\mu_1^2},\]
provided
\[\mu_1^2\le \frac 14 (\mu_2^2+2\sqrt{1+\mu_2^2}-2),\]
and check that this is consistent with the condition $0<\mu_1<\mu_2$
made before. In order to make a specific choice, we assume this
inequality for $\mu_1$ and hence obtain that
\begin{equation}
  \label{eq:choiceepsilon}
  \eps=\sqrt{1+\mu_1^2}-1.
\end{equation}
Now, in order to make the energy dissipation matrix positive, we must
choose $j$ so that for all $x\in U$,
$$
\aligned
\alphat_1(x) & :=\alpha_1(x)+(j-2)\kappa\eps>1,
\\
\alphat_2(x) & :=\alpha_2(x)+(j-2)\kappa\eps> 1 - 2k(x);
\endaligned
$$
cf.\ \eqref{eq:increasealpha}. These two inequalities are
satisfied for all functions $k$ under our assumptions if in particular
\begin{equation}
  \label{eq:estimatej}
  j>2+\frac{\sqrt{4(\Delta k)^2+\mu_2^2}}
  {\kappa(\sqrt{1+\mu_1^2}-1)}.
\end{equation}
In any case, we choose the maximal value for $\mu_1$
\begin{equation}
  \label{eq:choicemu1}
  \mu_1:=\frac 12 \sqrt{\mu_2^2+2\sqrt{1+\mu_2^2}-2}, 
\end{equation}
since this minimizes the value on the right side of
\eqref{eq:estimatej}.  We find that for this value of $\mu_1$, the
right side of \eqref{eq:estimatej} is monotonically decreasing in
$\mu_2$ and diverges to $+\infty$ for $\mu_2\to 0$ for all
values of $\Delta k$.

\begin{theorem}[Well-posedness theory for the Gowdy equations] 
  \label{th:well-posedness}
Consider some asymptotic data
$$
P_*=-k, \, P_{**}, \, Q_*,  \, Q_{**} \in C^\infty(U),
$$
where $k$ is a smooth function $U\to
  (1/2-\Delta k,1/2+\Delta k)$ for a constant $\Delta k\in
  (0,1/2)$. Then, for the Gowdy equations with these prescribed data are well-posed in certain weighted Sobolev spaces. 
  
  More precisely, 
  the singular initial value problem with asymptotic
  solutions of order $j$  has a unique solution
  with remainder $w\in X_{\delta,\alpha+(j-2)\kappa\eps,\infty}$
  for some sufficiently small $\delta>0$ and some $\kappa<1$. Here, 
 the exponents $\alpha=(\alpha_1,\alpha_2)$ and $\eps$ are given in 
  \Eqsref{eq:choicealpha1}, \eqref{eq:choicealpha2}, and
  \eqref{eq:choiceepsilon} explicitly in terms of the data and parameters $\mu_1, \mu_2$ chosen such that  
   $\mu_1$ is an explicit expression in $\mu_2$ given in  
  \Eqref{eq:choicemu1} while
   $\mu_2$ is a sufficiently close to (but smaller than) $\sqrt{1-4(\Delta k)^2}$, and the order of differentiation $j$ satisfies 
$$
j > 2+\frac{2}{\sqrt{3-4(\Delta k)^2+2\sqrt{2-4(\Delta k)^2}}-2}.
$$
\end{theorem}

The above condition implies that to reach $\Delta k\to 0$ we need $j>7$, while
$\Delta k\to 1/2$ requires $j\to\infty$. Although our
estimates may not be quite optimal, the latter implication cannot be avoided.


\subsection{Fuchsian analysis for the function 
  \texorpdfstring{$\Lambda$}{Lambda}}
\label{sec:Lambda}

So far we have considered the equations \Eqref{eq:originalgowdy} for
$P$ and $Q$. We can henceforth assume that these equations are solved
identically for all $t>0$ (and $t\le\delta$ for some $\delta>0$) and
that hence $P$ and $Q$ are given functions with leading-order behavior
\Eqref{eq:leadingorderPQ} and remainders in a given
$X_{\delta,\alpha,k}$. The equations which remain to be solved in
order to obtain a solution of the full Einstein's field equations are
\Eqref{eq:evolLambda} and \Eqref{eq:constraints}. In particular we are
interested in the function $\Lambda$ in order to obtain the full
geometrical information. We must compute $\Lambda$ also as a singular
initial value problem with ``data'' on the singularity analogously to
$P$ and $Q$. The following discussion resembles the previous one and
we only discuss new aspects now.

Clearly, the three remaining equations are overdetermined for
$\Lambda$ and hence solutions will exist only under certain
conditions. Let us define the following ``constraint quantities'' from
\Eqref{eq:constraints}
\begin{gather*}
  C_1(t,x):=-\del_t\Lambda+t(P_x)^2+e^{2P}t(Q_x)^2+t(\del_tP)^2
  +e^{2P}t(\del_tQ)^2,\\
  C_2(t,x):=-\Lambda_x+2P_x DP+2e^{2P}Q_x DQ.
\end{gather*}
Moreover, we define
\[H(t,x):=-\Lambda_{tt}+\Lambda_{xx}+ P_x^2-P_t^2+e^{2P}(Q_x^2-Q_t^2)
\]
from \Eqref{eq:evolLambda}. From the
evolution equations for $P$ and $Q$, we find
\begin{equation}
  \label{eq:constrprop}
  \del_t C_1=\del_x C_2+H,\quad \del_t C_2=\del_x C_1.
\end{equation}
These equations have the following consequences. Suppose that we use
\Eqref{eq:constraints2} as an evolution equation for $\Lambda$. This
implies that $C_1\equiv 0$ for all $t>0$. Moreover, suppose that we
prescribe data at some $t_0>0$ (indeed $t_0$ is allowed to be zero
later) so that $C_2(t_0,x)=0$ for all $x\in U$. Then the equations
imply that $H\equiv 0$ and $C_2\equiv 0$ for all $t>0$ and thus we
have constructed a solution of the full set of field
equations. Alternatively, let us use \Eqref{eq:evolLambda} as the
evolution equation for $\Lambda$, i.e.\ $H\equiv0$. Suppose that we
prescribe data so that $C_1(t_0,x)=C_2(t_0,x)=0$ at some $t_0$. It
follows that $C_1\equiv C_2\equiv 0$ for all $t>0$ because the
evolution system \Eqref{eq:constrprop} for $C_1$ and $C_2$ is
symmetric hyperbolic. Again, Einstein's field equations are
solved. 

Now, we want to consider the case $t_0=0$. First note that
\Eqsref{eq:constrprop} is regular even at $t=0$.  Suppose that $P$ and
$Q$ are functions with leading-order behavior
\Eqref{eq:leadingorderPQ} and remainders in a given
$X_{\delta,\alpha,k}$ with $k\ge 1$. If there exists a function $w_3$
so that
\begin{equation}
  \label{eq:expansionLambda}
  \Lambda(t,x)=\Lambda_*(x)\ln t+\Lambda_{**}(x)+w_3(t,x)
\end{equation}
with $w_3$ converging to zero in a suitable norm at $t=0$ and
\begin{equation}
  \label{eq:asymptconstr}
  \Lambda_*(x)=k^2(x),\quad
  \Lambda_{**}(x)=\Lambda_0
  +2\int_0^x k(\tilde x)(-\del_{\tilde x}P_{**}(\tilde x) +2
  e^{2P_{**}(\tilde x)}Q_{**}(\tilde x)\del_{\tilde x} Q_*(\tilde x))\,d\tilde x,
\end{equation}
where $\Lambda_0$ is an arbitrary real constant,
then 
\[\lim_{t\rightarrow t_0} C_2=0.\]
Let us first use \Eqref{eq:constraints2} as an evolution equation for
$\Lambda$. One can show easily that there exists a unique solution for
$\Lambda$ for $t>0$ which obeys the two-term expansion above. Our
discussion before implies that \Eqref{eq:evolLambda} is solved
identically for all $t>0$. Hence we obtain a solution of the full
Einstein's field equation. Alternative, choose \Eqref{eq:evolLambda}
as the evolution equation for $\Lambda$ now. This equation can be
written in second-order hyperbolic Fuchsian form
\[D^2\Lambda-t^2\del_x^2\Lambda
=(t\del_xP)^2+(D\Lambda-(DP)^2)+e^{2P}((t\del_x
  Q)^2-(DQ)^2).
\] 
Indeed this equation is compatible with the leading-order expansion
\Eqref{eq:expansionLambda} at $t=0$ and we can show well-posed of this
singular initial value problem in the same way as we did for the
functions $P$ and $Q$ before using the results of
\cite{BeyerLeFloch1}. In particular, for any asymptotic data
$\Lambda_{*}$ and $\Lambda_{**}$, there exists a unique solution of
this equation $\Lambda$ with remainder $w_3$ in a certain space
$X_{\delta,\alpha,k}$. Uniqueness implies that the solution $\Lambda$
of this equation coincides with the solution for $\Lambda$ obtained
using \Eqref{eq:constraints2} as the evolution equation. Hence, we
have $C_1\equiv C_2\equiv 0$ for all $t>0$, and thus also this method
yields a solution of the full Einstein's field equations.

Note that periodicity implies that the asymptotic data for $P$ and $Q$
must satisfy the relation
\[\int_0^{2\pi} k(\tilde x)(-\del_{\tilde x}P_{**}(\tilde x) +2
e^{2P_{**}(\tilde x)}Q_{**}(\tilde x)\del_{\tilde x} Q_*(\tilde
x))\,d\tilde x=0
\]
in the case of smooth solutions.


\section{Numerical solutions of the singular initial value problem}
\label{sec:numericalpart}

\subsection{The Fuchsian numerical algorithm}

We proceed now with the numerical approximation of the singular
initial value problem associate with second-order hyperbolic Fuchsian equations.
The approximation algorithm proposed now is motivated by our proof of
\Propref{P1-prop:existenceSVIPlinear} in \cite{BeyerLeFloch1}. 
For
linear source-terms, at least, we have shown that the solution of the 
singular initial value problem can be approximated by  
solutions to the \textit{regular initial value problem}. The regular
initial value problem is defined by data not at the singular time
$t=0$, but rather at some $t_0>0$, and by considering the evolution toward the future
(i.e.~away from the singular time $t=0$). Then, letting $t_0\to 0$, 
the sequence of solutions to these regular problems, referred to as 
\textit{approximate solutions},
converges toward the solution of the singular initial
value problem.  In \cite{BeyerLeFloch1}, we established 
an explicit error estimate for these approximate solutions and the result should extend 
to nonlinear source-terms satisfying a Lipschitz continuity conditions.

The regular initial value problem for second-order hyperbolic
equations corresponds to the standard initial value problem for a
system of (non-linear) wave equations, and there exists a huge amount
of numerical techniques for computing solutions. However, a
second-order Fuchsian equation written out with the standard
time-derivative $\del_t$ (instead of $D$) clearly involves factors
$1/t$ or $1/t^2$. Although these are finite for the regular initial
value problem, they still can cause severe numerical problems when the
initial time $t_0$ approaches zero, due to the finite representation of
numbers in a computer. In order to solve this problem, we introduce a
new time coordinate 
$$
\tau:=\ln t, 
$$
 and observe that
$D=\del_\tau$. For instance, the following Euler-Poisson-Darboux equation
was already treated in 
\cite{BeyerLeFloch1} 
\begin{equation}
  \label{eq:EPD}
  \del_\tau^2v-\lambda\, \del_\tau v-e^{2\tau}\del_x^2 v=0,
\end{equation}
where $v$ is the unknown and $\lambda$ is assumed to be a non-negative constant.  Therefore, 
there is no singular term in this equation; the main price which we
pay, however, is that the singularity $t=0$ has been ``shifted to''
$\tau=-\infty$. Another disadvantage is that the characteristic speed
of this equation (defined with respect to the $\tau$-coordinate) is
$e^\tau$ and hence increases exponentially with time.  For any
explicit discretization scheme, we can thus expect that the
CFL-condition will always be violated when we evolve into the future
from some time on. We must either adapt the time step to the
increasing characteristic speeds, or, when we decide to work with
fixed time resolution, accept the fact that (for any given
resolution) the numerical solution will eventually become
instable. However, this is not expected to be a severe problem, since
one can compute the numerical solution with respect to the
$\tau$-variable until some finite positive time when the numerical
solution is still stable and if necessary switch to a discretization
scheme based on the original $t$-variable afterwards. All the
numerical solutions presented in this paper are obtained with respect
to the $\tau$-variable without adaption.

We can simplify the following discussion slightly by writing (and
implementing numerically) the equation not for the function $v$ but
for the remainder $w=v-u_0$ with $u_0$ being the canonical two-term
expansion defined in \Eqref{P1-eq:SIVPDefV} in \cite{BeyerLeFloch1}
which is determined by given asymptotic data. According to the proof
of \Propref{P1-prop:existenceSVIPlinear} in \cite{BeyerLeFloch1}, the
remainder $w$ of any approximate solution is determined by initial
data
\[w(\tau_0,x)=0,\quad \del_\tau w(\tau_0,x)=0,\qquad x\in U,
\] 
for some $\tau_0\in\R$ successively going to $-\infty$.

Inspired by Kreiss et al.\ in \cite{Kreiss} and by the
general idea of the ``method of lines'', we proceed as follows to
discretize the equation. First we consider second-order Fuchsian
ordinary differential equations (written for scalar equations for simplicity)
$$
\del_\tau^2 w+2 a\,\del_\tau w+b\, w=f(\tau),
$$ 
where $f$ is a given function and the coefficients $a$ and $b$ are
constants. We discretize the time variable $\tau$ so that
$\tau_n:=\tau_0+n \Delta\tau$, $w_n:=w(\tau_n)$ and $f_n:=f(\tau_n)$
for some time step $\Delta\tau>0$. Then the equation is discretized in
second-order accuracy as
\begin{equation}
  \label{eq:numericalschemeODE}
  \frac{w_{n+1}-2w_n+w_{n-1}}{(\Delta\tau)^2}
  +2a \frac{w_{n+1}-w_{n-1}}{2\Delta\tau}+b w_n=f_n.
\end{equation}
Solving this for $w_{n+1}$ allows to compute the solution $w$ at the
time $\tau_{n+1}$ from the solution at the given and previous time
$\tau_n$ and $\tau_{n-1}$, respectively. At the initial two time steps
$\tau_0$ and $\tau_1$, we set, consistently with the initial data for
$w$ at $\tau_0$ above,
\begin{equation}
  \label{eq:initializescheme}
  w_0=0,\quad w_1=\frac 12 (\Delta\tau)^2 f(\tau_0).
\end{equation}
We will refer to this scheme as the \textit{Fuchsian ODE solver}.

The idea of the ``method of lines'' for Fuchsian partial differential
equations is to discretize also the spatial domain with the spatial
grid spacing $\Delta x$ and to use our Fuchsian ODE solver to
integrate one step forward in time at each spatial grid point. The
source-term function $f$, which might now depend on the unknown itself
and its first derivatives, is then computed from the data on the
current or the previous time levels. Here we understand that spatial
derivatives in the source-term are discretized by means of the
second-order centered stencil using periodic boundary conditions.  A
problem is that $f$, besides spatial derivatives, can also involve
time derivatives of the unknown $w$ (in fact this can be the case for
Fuchsian ordinary differential equations when the source term depends
on the time derivative of the unknown). In order to compute those time
derivatives in a systematic manner in second-order accuracy,
 i.e.~without changing the stencil of the Fuchsian ODE solver, we made the
following choice. In the code we store the numerical solution not only
on two time levels, as it is necessary up to now for the scheme given
by \Eqref{eq:numericalschemeODE} and \eqref{eq:initializescheme}, but
on a further third past time level. The time derivatives in the
source-term can then be computed from data at the present and previous
time steps only as follows
\[\del_{\tau} w(\tau_n)=\frac{3 w_n-4 w_{n-1}+w_{n-2}}{2\Delta\tau}
+O((\Delta\tau)^2).
\]
For this, we need to initialize three time levels and hence we set
\[w_2=2 (\Delta\tau)^2 f(\tau_0),\]
in addition to \Eqref{eq:initializescheme}.


\subsection{Euler-Poisson-Darboux equation}

We first tested the Fuchsian ODE solver on Fuchsian ordinary
differential equations. However, in the following, we consider 
the P.D.E.'s set-up directly and present numerical results for the
Euler-Poisson-Darboux equation \eqref{eq:EPD}. The reason for
considering this equation is that can be considered as a linear
version of each of the Gowdy equations. Recall from
\cite{BeyerLeFloch1} that the singular initial value problem with
two-term asymptotic data for this equation is well-posed for $0\le
\lambda<2$ and becomes ill-posed for $\lambda=2$. We study now this
singular initial value problem numerically for $\lambda>0$, i.e.~we
look for solutions
\[v(t,x)=u_*(x)+u_{**}(x)t^\lambda+w(t,x),
\] 
with remainder $w$. We choose the asymptotic data $u_*=\cos x$,
$u_{**}=0$. Note that in this case, this leading-order behavior is
consistent even with the case $\lambda=0$. But according to the
discussion in \cite{BeyerLeFloch1}, it is not consistent with
$\lambda=2$, and we expect that this becomes visible in the numerical
solutions. For $u_{**}=0$ and $0<\lambda<2$, we can show that the
leading-order behavior of the remainder at $t=0$ is
\begin{equation}
  \label{eq:leadingorderexact}
  w(t,x)=u_*(x)\left(-\frac 1{2(2-\lambda)}t^2 +\frac
  1{8(2-\lambda)(4-\lambda)}t^4+\ldots\right).
\end{equation}

\begin{figure}[ht]
 \subfigure[$\lambda=1.99$.]{\includegraphics[width=0.49\linewidth]{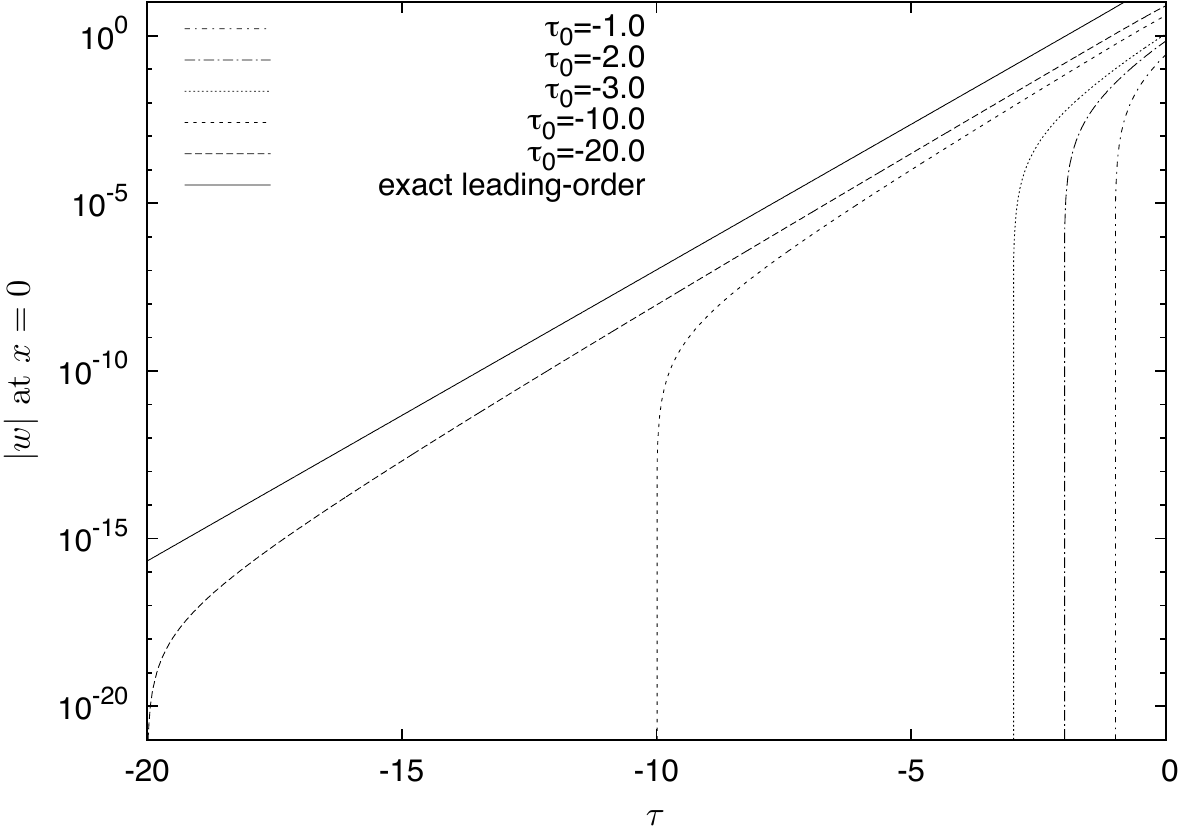}}
  \subfigure[$\lambda=1.90$.]{\includegraphics[width=0.49\linewidth]{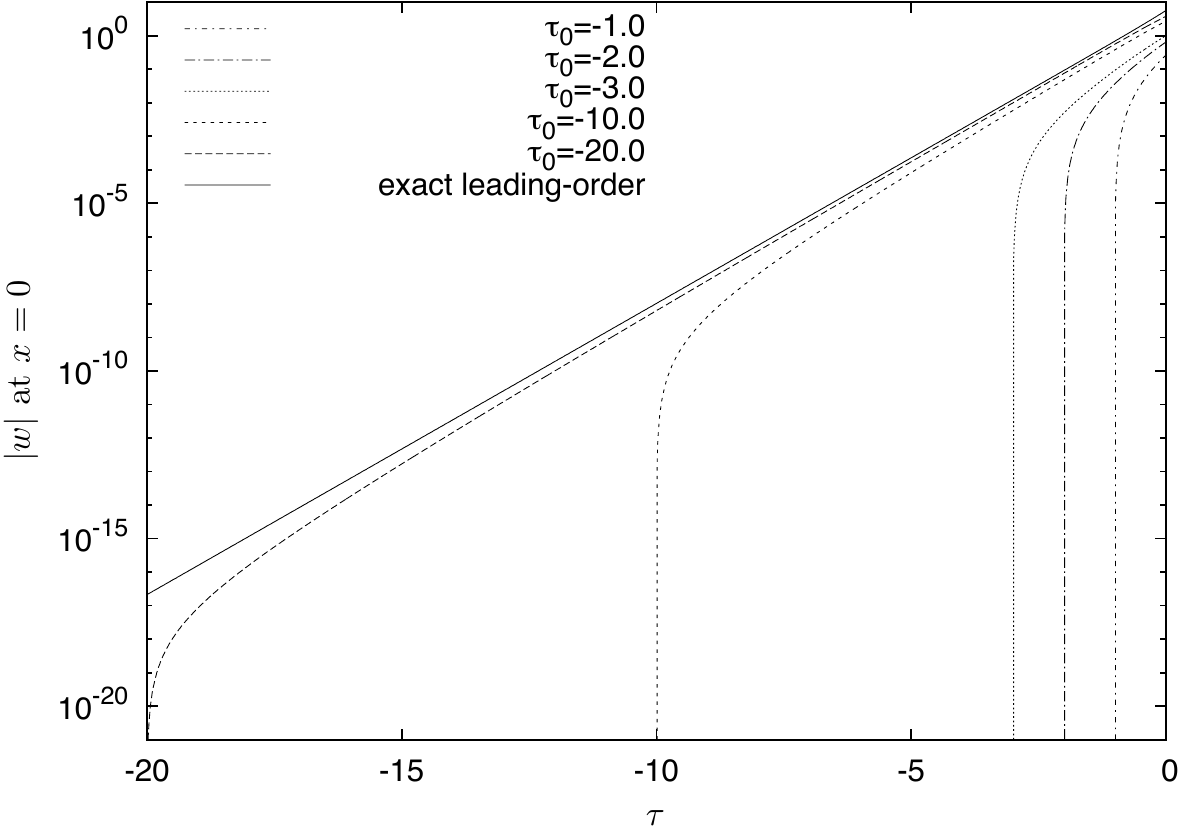}} \\
 \subfigure[$\lambda=1.00$.]{\includegraphics[width=0.49\linewidth]{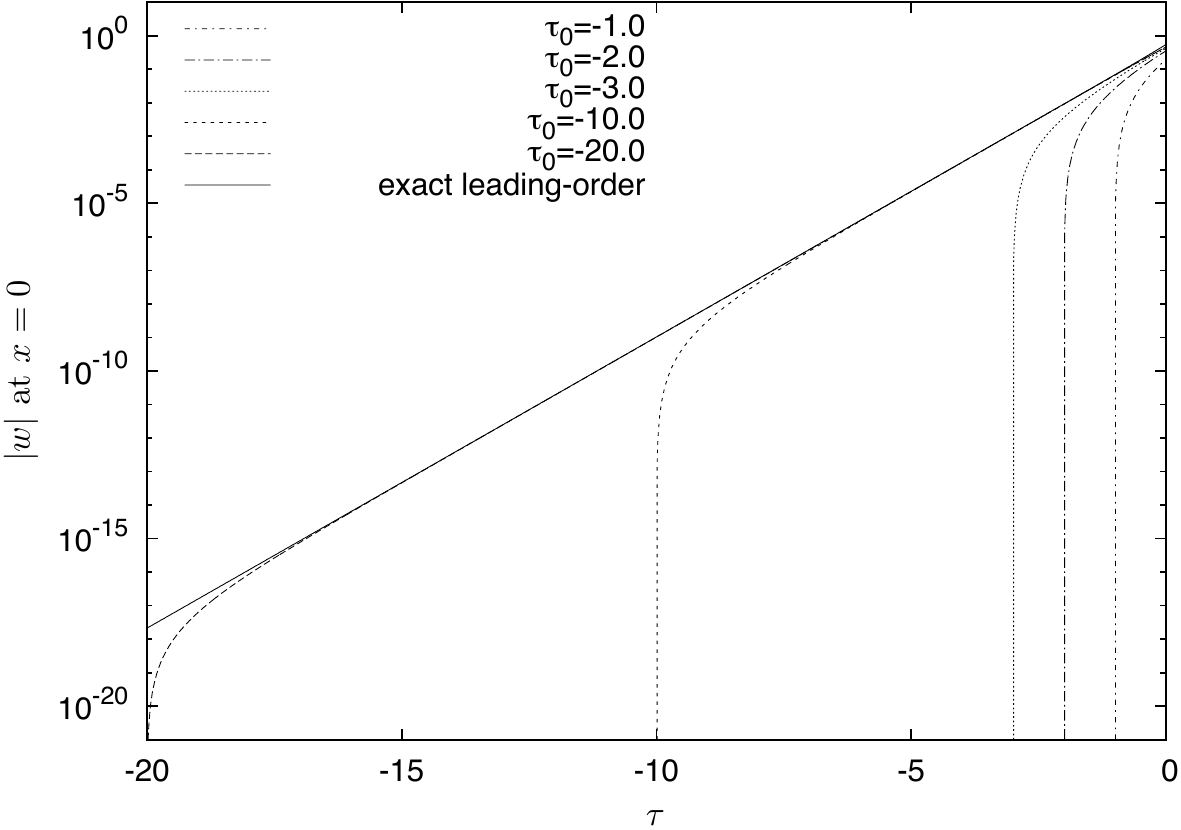}}
 \subfigure[$\lambda=0.01$.]{\includegraphics[width=0.49\linewidth]{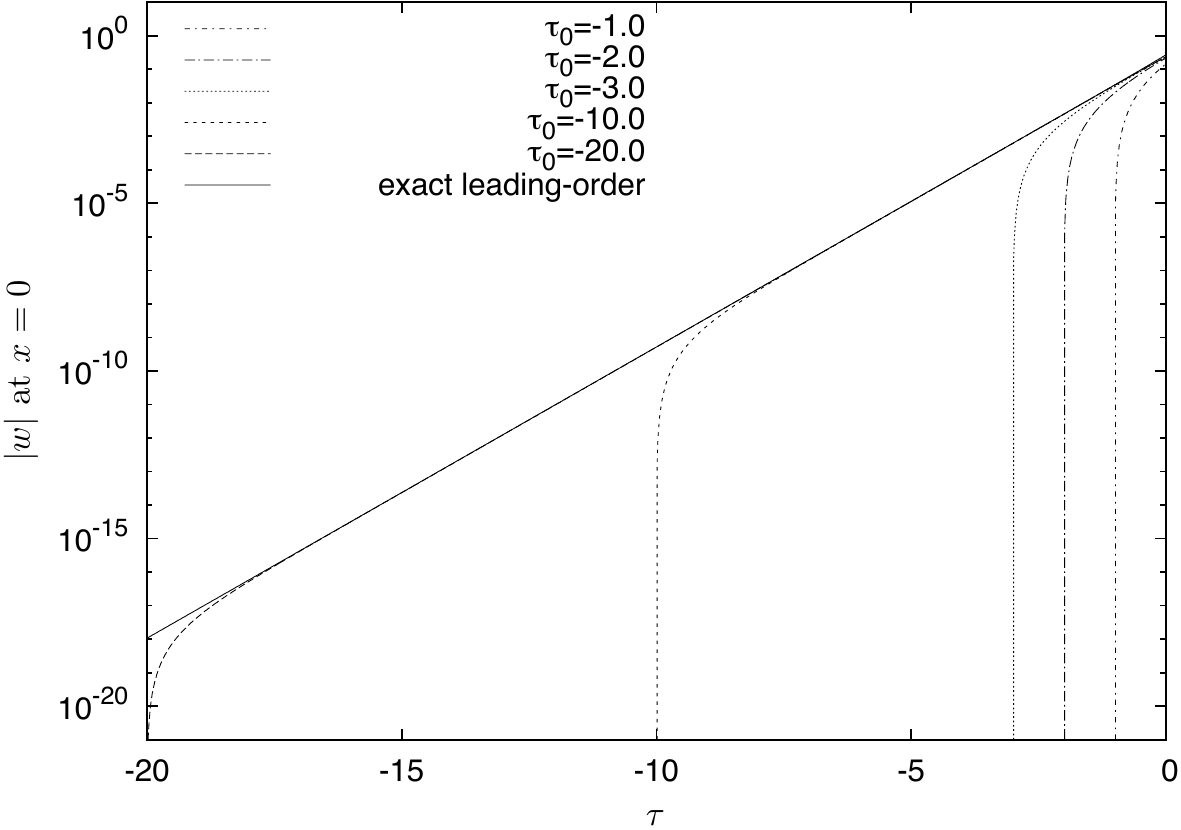}}
\caption{Euler-Poisson-Darboux equation.}
\protect \label{fig:EPDNumerics}
\end{figure}

First we check that the numerical solutions converge in second-order
when $\Delta\tau$ and $\Delta x$ are changed proportionally to each
other for a given choice of initial time $\tau_0>0$. We do not discuss
this further here, but eventually we choose the resolution so that
discretization errors are negligible in the following discussion; the
same is true for round-off errors, as we come back to later. In
\Figref{fig:EPDNumerics}, we show the following results obtained with
$N=20$, $\Delta\tau=0.003$. Here, $N$ is the number of spatial grid
points, i.e.~one has $\Delta x=2\pi/N$. We find that for this CFL-parameter
$\Delta t/\Delta x\approx 0.01$, the runs are stable until
$\tau\approx 5$. For each of the plots of \Figref{fig:EPDNumerics}, we
fix a value of $\lambda$ and study the convergence of the approximate
solutions to the (leading-order of the) exact solution
\Eqref{eq:leadingorderexact} for various values of the initial time
$\tau_0$. We plot the value at one spatial point $x=0$ only.  The
convergence rate for $\tau_0\to-\infty$ is fast if $\lambda=1$
or $\lambda=0.01$, but becomes lower, the more $\lambda$ approaches
the value $2$, where it becomes zero.  This is in exact agreement with
our expectations and consistent with the error estimates derived in
\cite{BeyerLeFloch1}.  Hence the numerical results are very promising.

To close the discussion of this test case, let us add some comment about
numerical round-off errors. All numerical runs in this paper were done
with double-precision (binary64 of IEEE 754-2008), where the real
numbers are accurate for $16$ decimal digits. However, for the case
$\tau_0=-20$ for instance, the second spatial derivative of the
unknown in the equation is multiplied by $\exp(-40)\approx 10^{-18}$
at the initial time which is not resolved numerically and hence could
possibly lead to a significant error. This, however, does not seem to
be the case since we obtained virtually the same numerical solution
with quadruple precision (binary128 of IEEE 754-2008), i.e.~when the
numbers in the computer are represented with $34$ significant decimal
digits.


\subsection{Singular initial value problem for the Gowdy equations}

We continue our discussion with the singular initial value problem
for the Gowdy equations. In all of what follows we
consider the singular initial value problem with two-term asymptotic
data for the Gowdy equations. The fact that this works
very well and we get good convergence can be seen as an indication that
this singular initial value problem is well-posed. Recall from
\Theoremref{th:well-posedness} that our analytical techniques are only
sufficient to show that the initial value problem with asymptotic
solutions of sufficiently high order is well-posed for the Gowdy equations.

\vskip.3cm 

\paragraph{\em Test~1. Homogeneous pseudo-polarized solutions.}
Before we proceed with ``interesting'' solutions of the Gowdy problem,
let us start with a test case for which we can construct an explicit
solution.  Let $\Pt$ and $\Qt$ be solutions of the polarized
equations in the homogeneous case, i.e.~one set  
$\Qt=0$ and $\Pt(t,x)=\Pt(t)$. In this case, it follows directly that the exact
solution of the Gowdy equations is
\[\Pt(t)=-k\ln t+\Pt_{**},\]
where both $k$ and $\Pt_{**}$ are arbitrary constants. By a reparametrization
of the Killing orbits of the form
\[\tilde x_2=x_2/\sqrt 2+x_3/\sqrt 2,\quad 
\tilde x_3=-x_2/\sqrt 2+x_3/\sqrt 2,
\] 
where $\tilde x_2$, and $\tilde x_3$ are the coordinates used to
represent the orbits of the polarized solution above, the same
solution gets reexpressed in terms of functions
\be
\label{hhhh}
P=\ln\cosh (-k\ln t+\Pt_{**}),\quad Q=\tanh (-k\ln t+\Pt_{**}).
\ee
Of course, these functions $(P,Q)$ are again solutions of
\eqref{eq:originalgowdy}. Asymptotically at $t=0$, they satisfy 
\begin{equation}
  \label{eq:exactpolarized}
  P=-k\ln t+(\Pt_{**}-\ln 2)+\ldots,\quad Q=1-2e^{-2\Pt_{**}}t^{2k}+\ldots,
\end{equation}
from which we can read off the corresponding asymptotic data. 

\begin{figure}[t]  
  \centering
  \subfigure[$k=0.5$.]{%
    \includegraphics[width=0.49\textwidth]{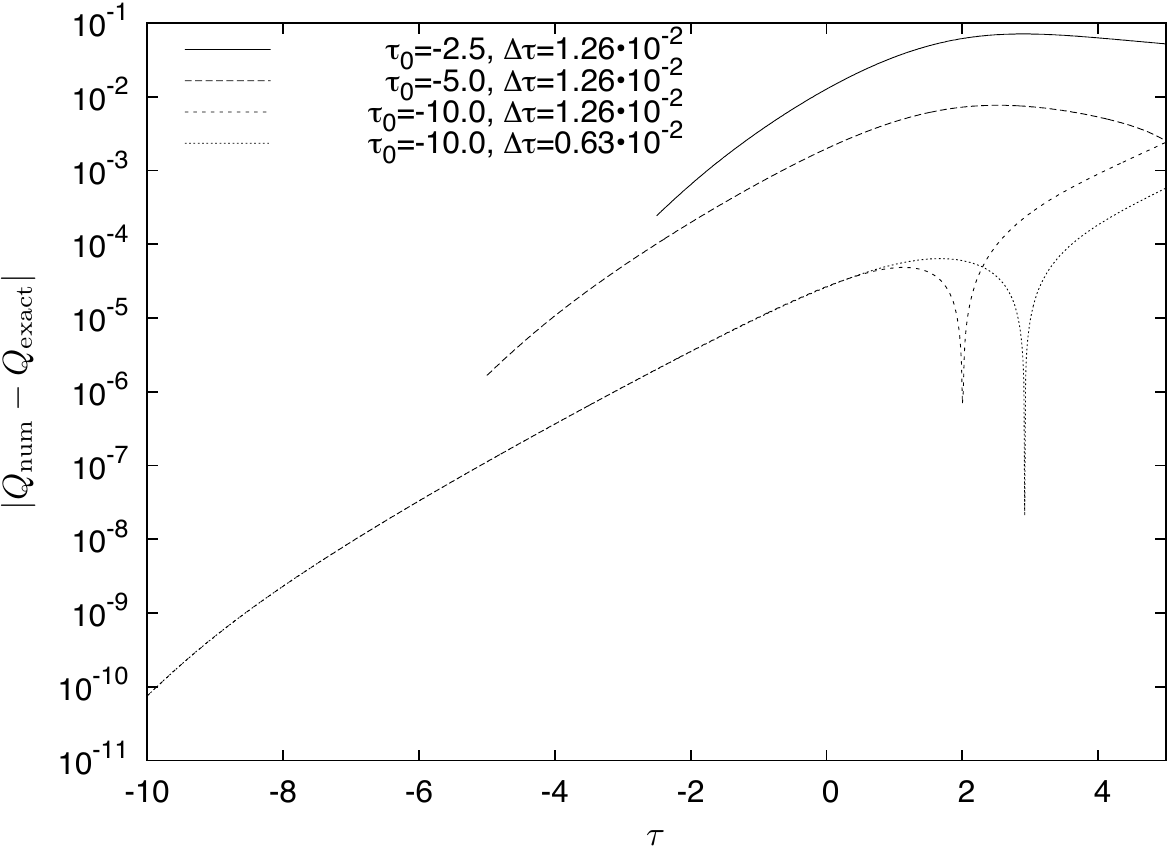}}
  \subfigure[$k=0.9$.]{%
    \includegraphics[width=0.49\textwidth]{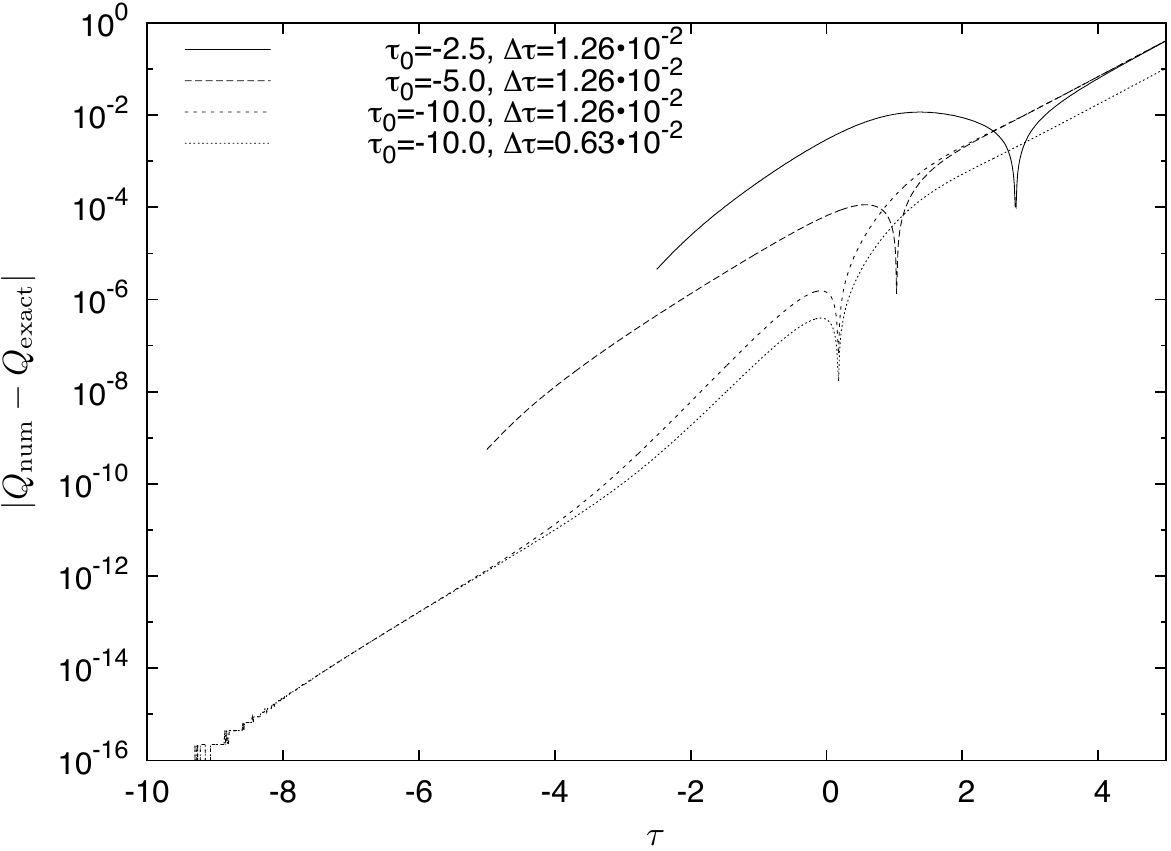}}
  \caption{Homogeneous pseudo-polarized case (Gowdy equations).}
  \label{fig:GowdyPseudoPolHomog}
\end{figure}
Now we compute the solutions corresponding to these asymptotic data
numerically and compare them to the exact solution
\Eqref{hhhh}. We pick $\Pt_{**}=1$, so that
$P_{**}=1-\ln 2$, $Q_{*}=1$ and $Q_{**}=-2 e^{-2}$. Since the solution
is spatially homogeneous -- in fact this is an ODE problem -- we only
need to do the comparison at one spatial point. The results are
presented in \Figref{fig:GowdyPseudoPolHomog} where we plot the
difference of the numerical and the exact value of $Q$ versus time for
various values of $\tau_0$. In the first plot, this is done for
$k=0.5$ and in the second plot for $k=0.9$. The plots confirm nice
convergence of the approximate solutions to the exact solution. The
fact that each approximate solution diverges from the exact solution
almost exponentially in time is a feature of the approximate solutions
themselves and not of the numerical discretization, as is checked by 
comparing two different values of $\Delta\tau$ in these plots.  From
our experience with the Euler-Poisson-Darboux equation, we could have
expected that the convergence rate is lower in the case $k=0.9$ than
in the case $k=0.5$ (note that $k$ plays the same role
$\lambda/2$). In the case of the Euler-Poisson-Darboux equation, the
rate of convergence decreases when $\lambda$ approaches $2$, due to
the influence of the second-spatial derivative term in the
equation. In the spatially homogeneous case here, however, this term
is zero and hence this phenomenon is not present. The ``spikes'' in
\Figref{fig:GowdyPseudoPolHomog} are just a consequence of the
logarithmic scale of the horizontal axes and the fact that the
numerical and exact solutions equal for one instance of time.

\vskip.3cm 

\paragraph{\em Test~2. General Gowdy equations}
\begin{figure}[t]  
  \centering
  \subfigure[$A=0.2$.]{%
    \includegraphics[width=0.49\textwidth]{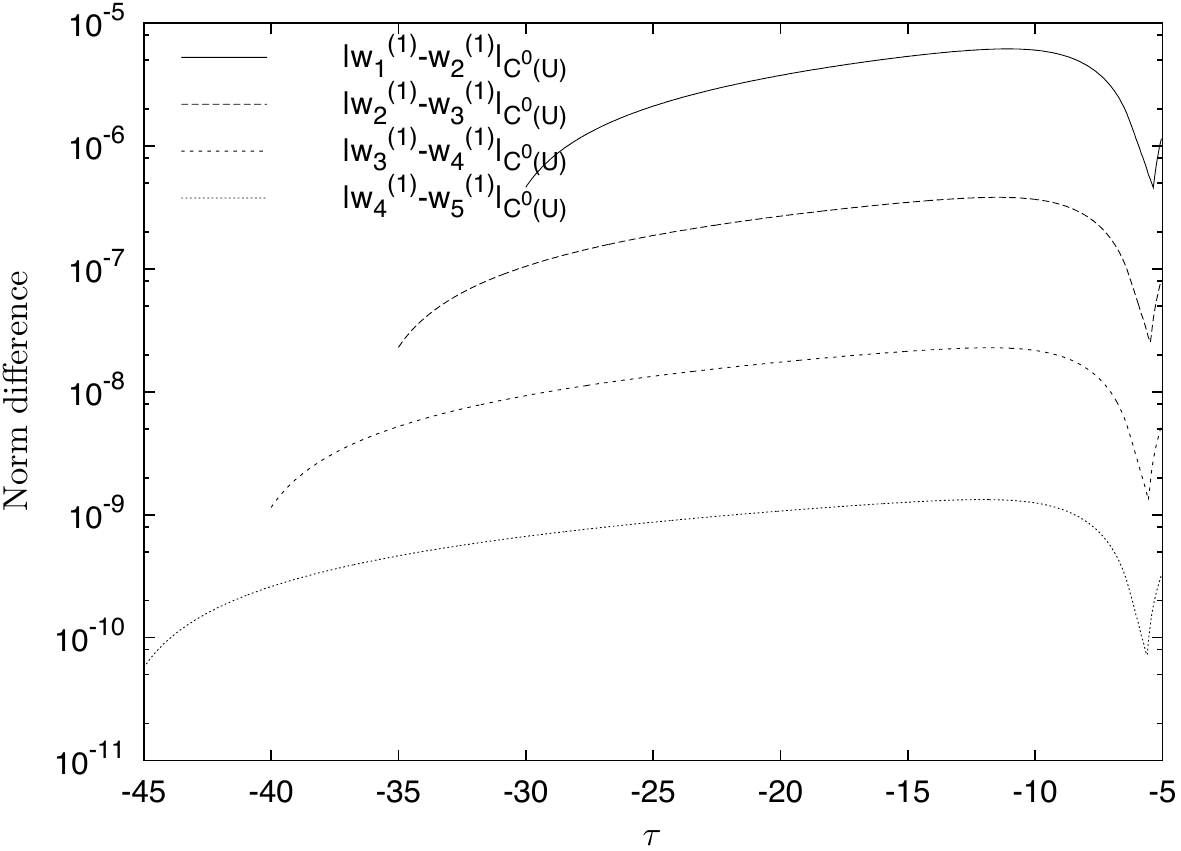}}
  \subfigure[$A=0.4$.]{%
    \includegraphics[width=0.49\textwidth]{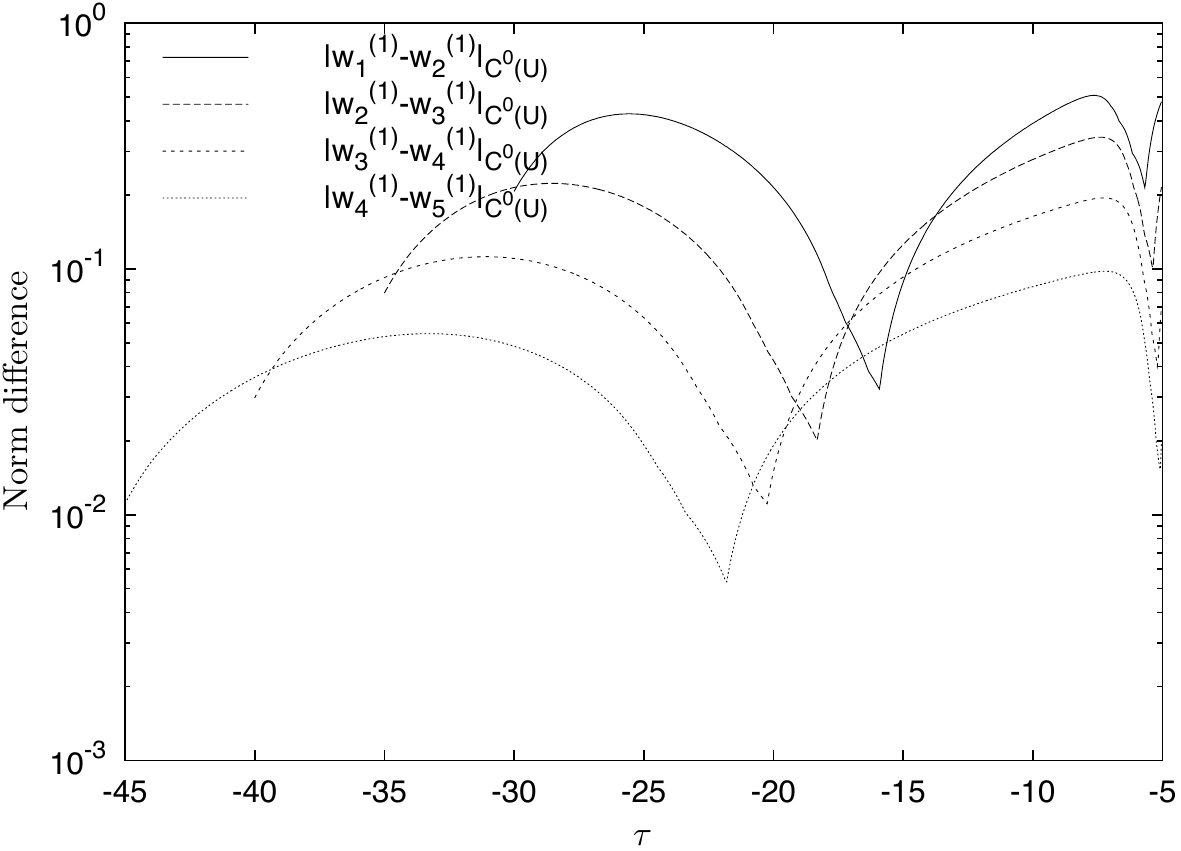}}
  \caption{Convergence for general Gowdy solutions.}
  \label{fig:GowdyGeneralConvergence}
\end{figure}
Now we want to study the convergence for a ``generic'' inhomogeneous
Gowdy case (still ignoring the equation for the quantity
$\Lambda$). Here we choose the following asymptotic data
$$
\aligned
& k(x)=1/2+A\cos(x),\qquad Q_*=1.0+\sin(x),
\\
& P_{**}=1-\ln 2+\cos(x),\qquad Q_{**}=-2 e^{-2},
\endaligned
$$  
with a constant $A\in (-1/2,1/2)$. We do not know of an explicit
solution in this case. In \Figref{fig:GowdyGeneralConvergence}, we
show the following numerical results for $A=0.2$ and $A=0.4$,
respectively. For the given value of $A$, we compute five approximate
solutions numerically with initial times $\tau_0=-30,-35,-40,-45,-50$
numerically, each with the same resolution $\Delta\tau=0.01$ and
$N=80$. The resolution parameters have been chosen so that the
numerical discretization errors are negligible in the plots of
\Figref{fig:GowdyGeneralConvergence}. Then, for each time step for
$\tau\ge-30$, we compute the supremum norm in space of the difference
of the remainders $w^{(1)}$ of the two approximate solutions given by
$\tau_0=-30$ and $\tau_0=-35$. In this way we obtain the first curve
in each of the plots of \Figref{fig:GowdyGeneralConvergence}. The same
is done for the difference between the cases $\tau_0=-35$ and
$\tau_0=-40$ for all $\tau\ge-35$ to obtain the second curve
etc. Hence these curves yield a measure of the convergence rate of the
approximation scheme, without referring to the exact solution.  In
agreement with our observation for the Euler-Poisson-Darboux equation,
the convergence rate is high if $k$ is close to $1/2$ and becomes
lower, the more $k$ touches the ``extreme'' values $k=0$ and $k=1$.

Much in the same way as for the Euler-Poisson-Darboux equation we find
that double precision is sufficient for these computations despite of
the fact that $\exp(2\tau)$ is $10^{-44}$ for $\tau=-50$.


\subsection{Gowdy spacetimes containing a Cauchy horizon}

The papers 
\cite{ChruscielIsenbergMoncrief,ChruscielIsenberg,ChruscielLake,HennigAnsorg,IsenbergMoncrief}
were devoted to the construction and characterization of Gowdy
solutions with Cauchy horizons, in particular in order to prove the
strong cosmic censorship conjecture in this class of spacetimes.
Spacetimes with Cauchy horizons are expected to have saddle and
physically ``undesired'' properties, in particular they often allow
various inequivalent smooth extensions, i.e.~the Cauchy problem of
Einstein's field equations does not select one of them uniquely.  Some
explicit examples are known, but most of the analysis is on the level
of existence proofs and asymptotic expansions.  

Hence, it is of 
interest to construct such solutions numerically and analyze
them in much greater detail than possible with purely analytic
methods. Constructing these solutions numerically, however, is 
delicate since the strong cosmic censorship conjecture suggests that
they are instable under generic perturbations. It can hence often be
expected that numerical errors would most likely ``destroy the Cauchy
horizon''. This is so, in particular, when the singular time at $t=0$ is
approached backwards in time from some regular Cauchy surface at
$t>0$. 

In the Gowdy case, however, there are clear criteria for the
asymptotic data so that the corresponding solution of the singular
initial value problem has a Cauchy horizon (or only pieces thereof; 
cf.~below) at $t=0$, as discussed in \cite{ChruscielIsenbergMoncrief}
for the polarized case and in \cite{ChruscielLake} for the general
case. Our novel method here allows us to construct such solutions with
arbitrary accuracy and it can hence be expected that this allows us to
study the saddle properties of such solutions.  Our main aim for the
following is to compute such a solution and with this demonstrate the
feasibility of our approach. A follow-up work will be devoted to the
numerical construction and detailed analysis of relevant classes of
such solutions.

Motivated by the results in \cite{ChruscielIsenbergMoncrief}, we choose
the asymptotic data as follows
\begin{align*}
  &k(x)=
  \begin{cases}
    1, & x\in [\pi,2\pi],\\
    1-e^{-1/x}e^{-1/(\pi-x)}, & x\in (0,\pi),
  \end{cases}&\quad &P_{**}(x)=1/2,\\
  &Q_*(x)=0,&\quad &Q_{**}(x)=\begin{cases}
    0, & x\in [\pi,2\pi],\\
    e^{-1/x}e^{-1/(\pi-x)}, & x\in (0,\pi),
  \end{cases}\\
  &\Lambda_*(x)=k^2(x),&\quad
  &\Lambda_{**}(x)=2.
\end{align*}
With these asymptotic data, the corresponding solution has a smooth
Cauchy horizon at $(t,x)\in \{0\}\times (\pi,2\pi)$ (namely where
$k\equiv 1$), and a curvature singularity at $(t,x)\in \{0\}\times
(0,\pi)$ (namely where $0<k<1$). Note that the function $k$ is smooth
everywhere (but not analytic). Our analysis
in \Sectionref{sec:wellposednessGowdy} shows that we are allowed to
set $k=1$ at some points since $\del_xQ_*=0$. This motivates our
choice of $Q_*$.  With this, our choice of $Q_{**}$ implies that the
solution is polarized on the ``domain of dependence''\footnote{The notion of
 ``domain of
  dependence'' for the singular initial value problem
follows easily from the energy estimate established in \cite{BeyerLeFloch1}.} of the
``initial data'' interval $(\pi,2\pi)$. All data were chosen
as simple as possible to be consistent with the constraints.

\begin{figure}[t]  
  \centering  
  \includegraphics[width=0.49\textwidth]{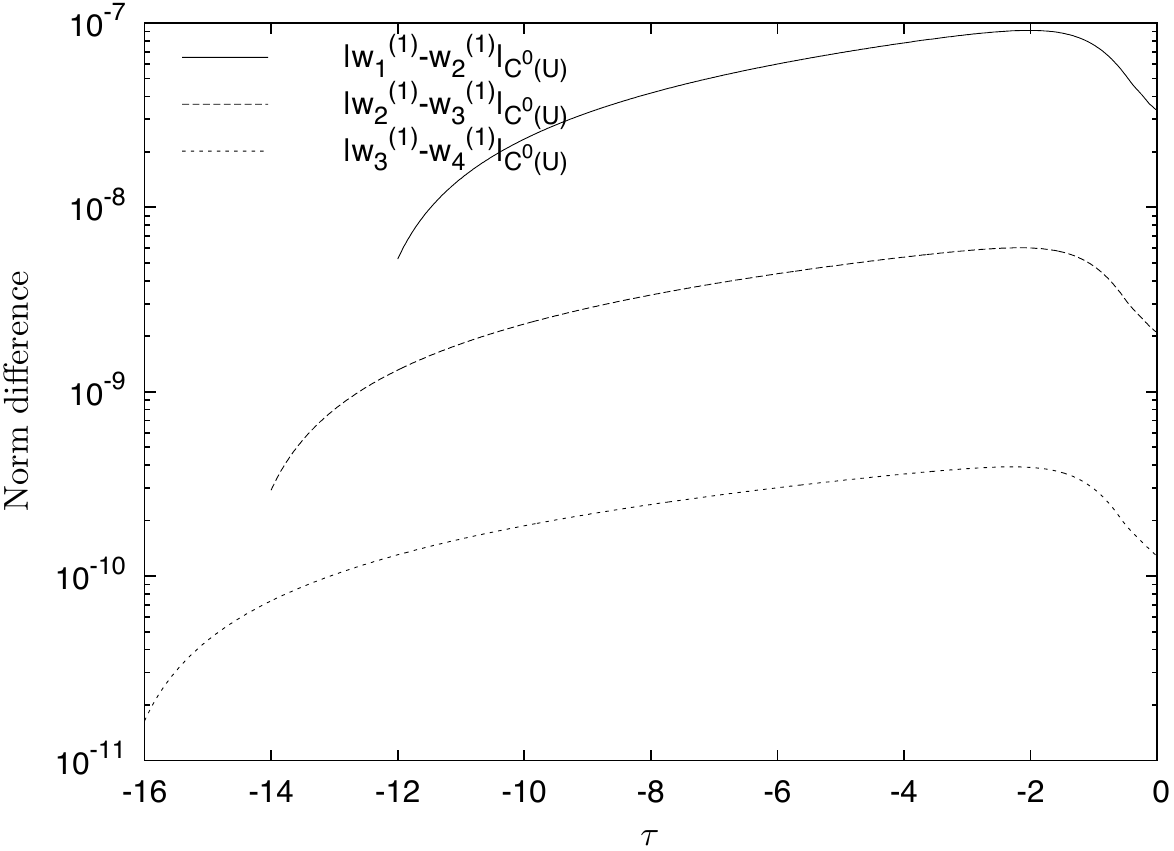}
  \caption{Convergence analysis for spacetimes with a Cauchy horizon.}
  \label{fig:CHConvergence}
\end{figure}
First we repeated the same error analysis as for the previous Gowdy
case, see~\Figref{fig:CHConvergence}. For all the runs in the plots,
we choose $N=500$, $\Delta\tau=0.005$ which guarantees that
discretization errors are negligible in the plot. We find that our
numerical method allows us to compute the Gowdy solution very
accurately. Here, we solve the full system for
$(P,Q,\Lambda)$.

\begin{figure}[t] 
  \centering
  \subfigure[Kretschmann at $\tau=-10.0$.]{%
    \includegraphics[width=0.49\textwidth]{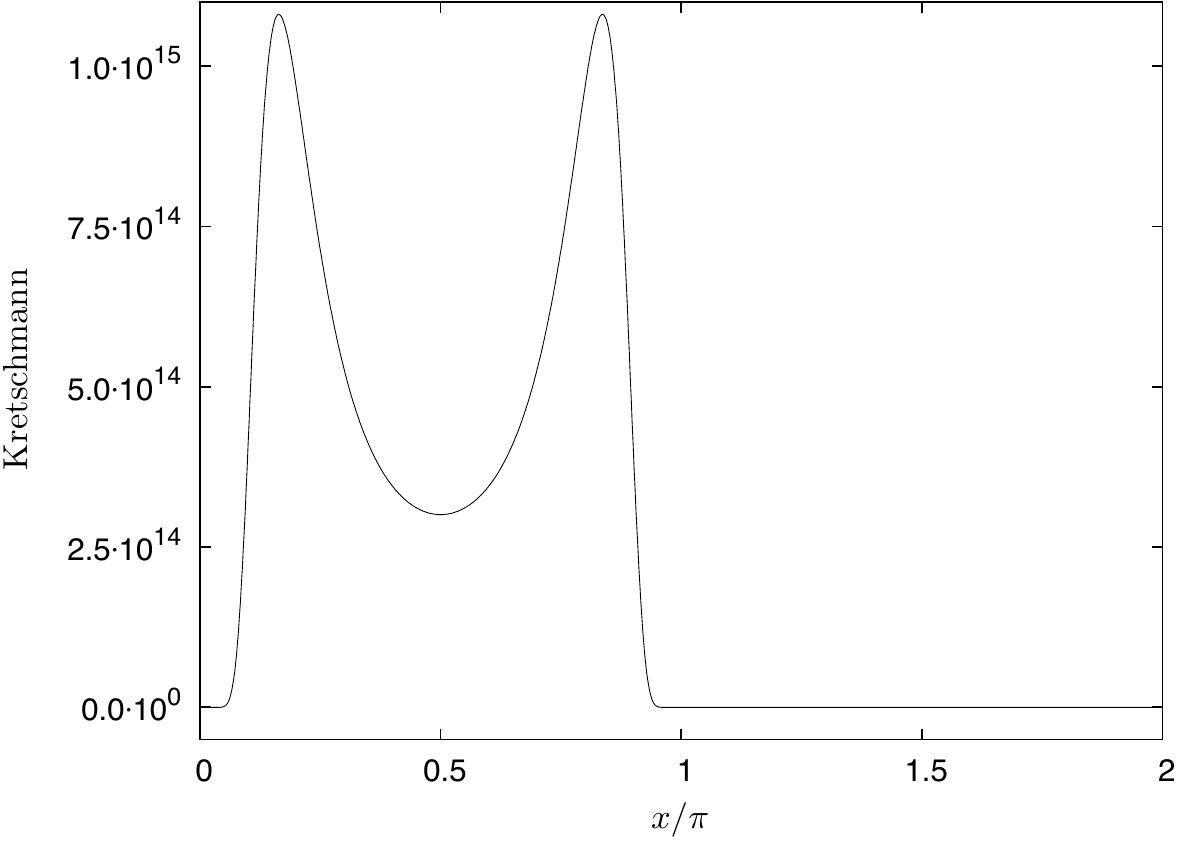}}
  \subfigure[Kretschmann at $\tau=0.0$.]{%
    \includegraphics[width=0.49\textwidth]{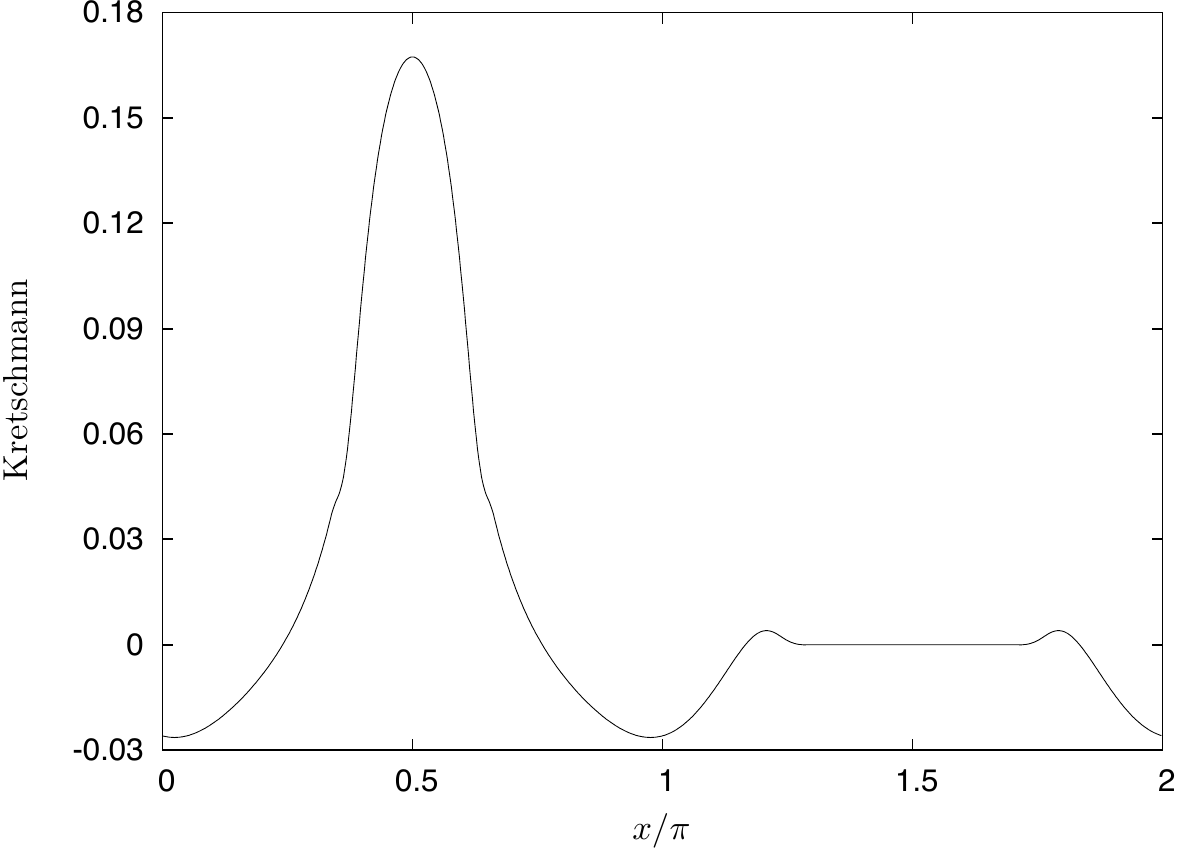}}

  \subfigure[Remainders of $P,Q$ at $\tau=-10.0$.]{%
    \includegraphics[width=0.49\textwidth]{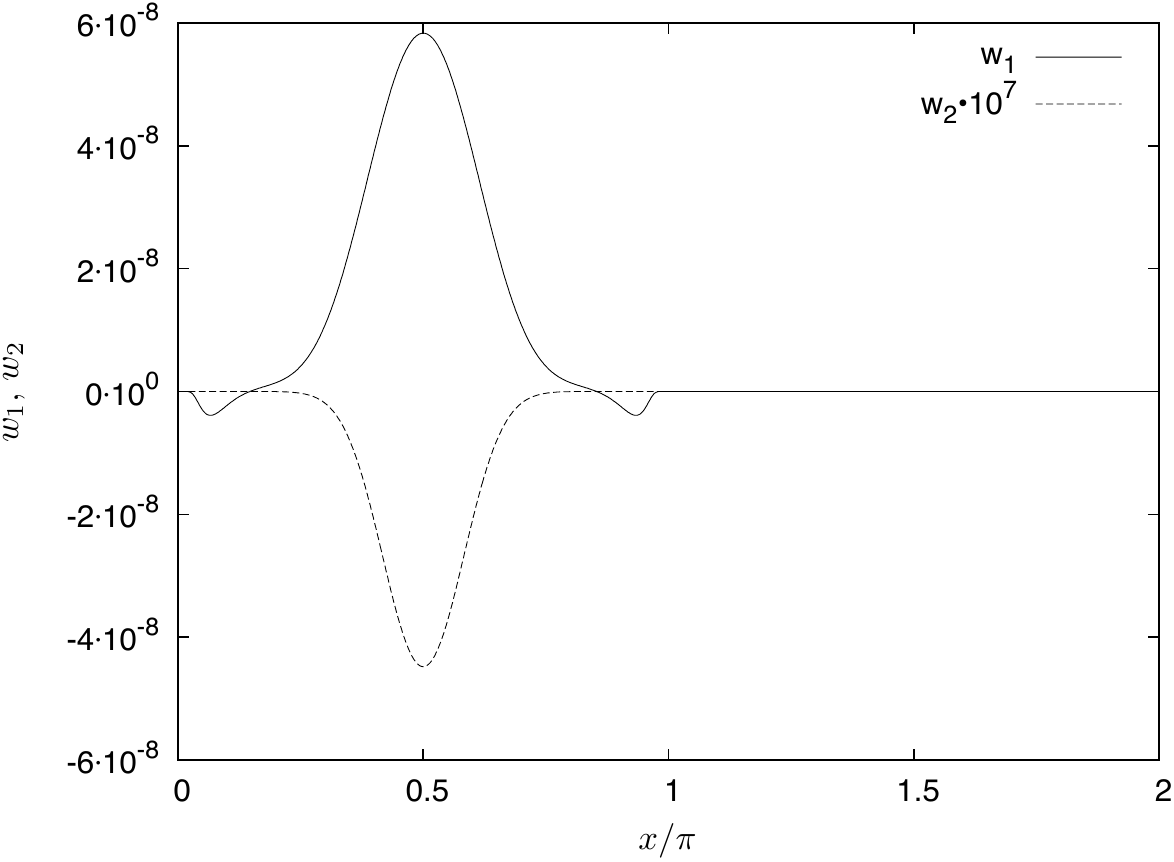}}
  \subfigure[Remainders of $P,Q$ at $\tau=0.0$.]{%
    \includegraphics[width=0.49\textwidth]{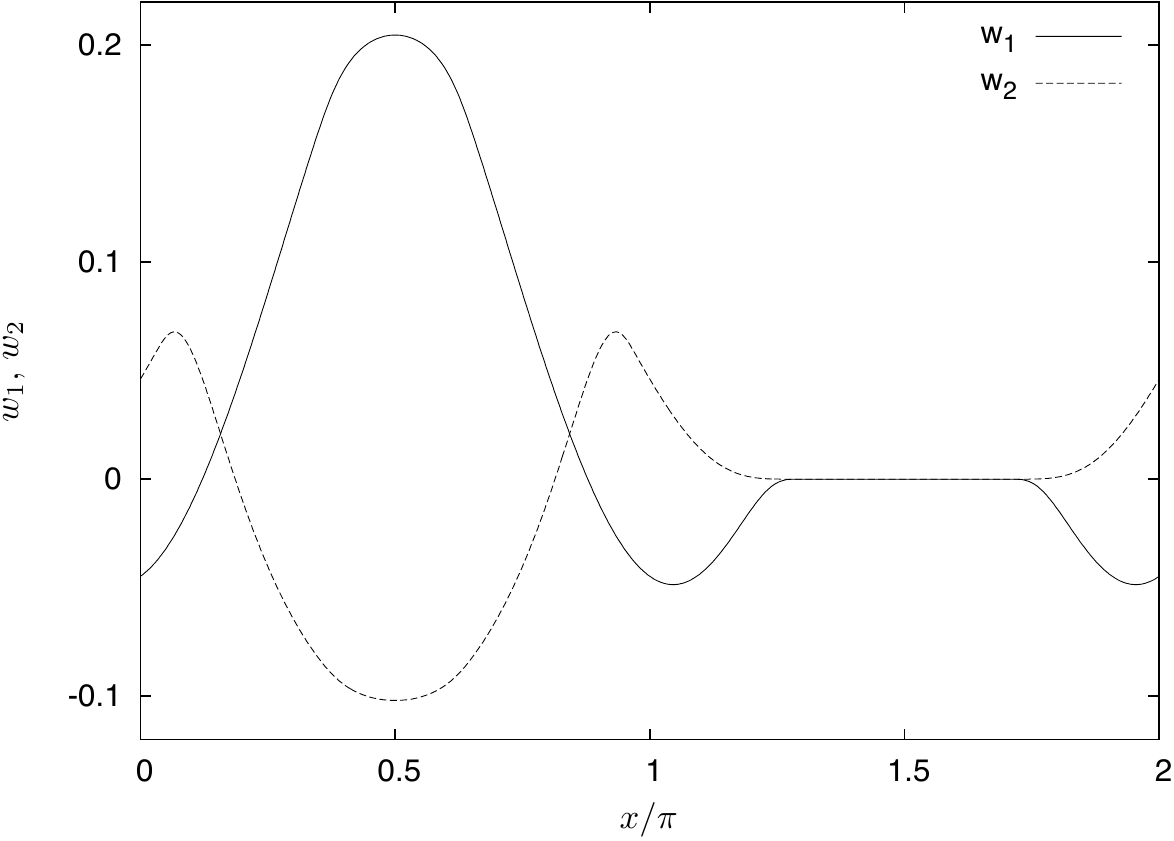}}
  \caption{Behavior of solutions with a Cauchy horizon.}
  \label{fig:CH1Dplots}
\end{figure}
In \Figref{fig:CH1Dplots}, we show the numerical
solution obtained from $N=1000$, $\Delta\tau=0.0025$ and $\tau_0=-18$.
We plot the Kretschmann scalar at two times $\tau=-10$ and
$\tau=0$. Hence, near the time $t=0$ (corresponding to
$\tau=-\infty$),
 the Kretschmann scalar is large on the
spatial interval $(0,\pi)$ while it stays bounded at $(\pi,2\pi)$. At
the later time, the curvature becomes smaller as expected. We also
plot the remainders $w^{(1)}$ and $w^{(2)}$ of $P$ and $Q$,
respectively. It is instructive to study how the polarized region inside
$(\pi,2\pi)$ gets ``displaced'' by the non-polarized solution. We refer the reader to \cite{BeyerLeFloch3}
for further investigations, especially 
a study of past-directed causal geodesics
approaching the $t=0$-hypersurface near the boundary point
$x=\pi$ at the intersection of the Cauchy horizon and the curvature
singularity. Furthermore, in \cite{BeyerLeFloch3} we will discuss trapped surfaces in a
neighborhood of $t=0$. 

 
\section*{Acknowledgements}

The authors were partially supported by the Agence Nationale de la
Recherche (ANR) through the grant 06-2-134423 entitled {\sl
  ``Mathematical Methods in General Relativity''} (MATH-GR). A 
first draft of this paper was written during the year 2008--2009 when
the first author (F.B.) was an ANR postdoctoral fellow at the Laboratoire J.-L. Lions.
 

\end{document}